\documentclass[showpacs,preprintnumbers,amsmath,amssymb,superscriptaddress,nofootinbib]{revtex4}
\usepackage{graphicx}
\usepackage{dcolumn}
\usepackage{bm}
\usepackage{longtable}
\usepackage[countmax]{subfloat}
\usepackage{epstopdf}  
\usepackage{easybmat}
\usepackage{soul}
\usepackage{longtable}
\usepackage{bm}
\usepackage{dcolumn}
\usepackage{epsfig}
\usepackage{graphicx}
\usepackage{subfigure}
\usepackage{comment}
\usepackage{amsfonts}
\usepackage{amsmath}
\usepackage{amssymb}
\usepackage{subfigure}
\usepackage{times}
\usepackage{appendix}
\usepackage{color}

\providecommand{\be}{\begin{equation}}
  \providecommand{\ee}{\end{equation}}
\providecommand{\bea}{\begin{eqnarray}}
  \providecommand{\eea}{\end{eqnarray}}
\providecommand{\ba}{\begin{eqnarray}}
  \providecommand{\ea}{\end{eqnarray}}
\providecommand{\beas}{\begin{eqnarray*}}
  \providecommand{\eeas}{\end{eqnarray*}}

\providecommand{\beni}{\begin{equation*}}
  \providecommand{\eeni}{\end{equation*}}

\def\s{\sigma}

\usepackage{epsfig}
\usepackage{amssymb}
\usepackage{amsmath}
\usepackage{amsthm}
\usepackage{latexsym}
\def\be{\begin{equation}}
\def\ee{\end{equation}}
\def\bc{\begin{center}}
\def\ec{\end{center}}

\begin{document}

\preprint{APS/124-QED}

\title{The exact Laplacian spectrum for the Dyson hierarchical network}

\author{Elena Agliari}
\affiliation{Dipartimento di Matematica, Sapienza Universit\`a di Roma, P.le A. Moro 2, 00185, Roma, Italy.}

\author{Flavia Tavani}
\affiliation{Dipartimento SBAI (Ingegneria), Sapienza Universit\`a di Roma, via A. Scarpa 16, 00161, Roma, Italy.}
\date{\today}

\begin{abstract} 
We consider the Dyson hierarchical graph $\mathcal{G}$, that is a weighted fully-connected graph, where the pattern of weights is ruled by the parameter $\sigma \in (1/2, 1]$. 
Exploiting the deterministic recursivity through which $\mathcal{G}$ is built, we are able to derive explicitly the whole set of the eigenvalues and the eigenvectors for its Laplacian matrix.
Given that the Laplacian operator is intrinsically implied in the analysis of dynamic processes (e.g., random walks) occurring on the graph, as well as in the investigation of the dynamical properties of connected structures themselves (e.g., vibrational structures and the relaxation modes), this result 
allows addressing analytically a large class of problems.
In particular, as examples of applications, we study the random walk and the continuous-time quantum walk embedded in $\mathcal{G}$, the relaxation times of a polymer whose structure is described by $\mathcal{G}$, and the community structure of $\mathcal{G}$ in terms of modularity measures.
\end{abstract} 

\maketitle

\section*{Introduction}

Most real-life networks display non-trivial features concerning their topological structure as well as the pattern of weights associated to their links. In fact, the arrangement of the connections between their constituent elements is typically neither purely regular nor purely random and, also, such connections are associated to weights accounting for wiring costs which may depend on the
distance (e.g., a physical
distance, a social distance) between elements to be linked \cite{Newman-2012, Gallos-PNAS2011, Serrano-PNAS2009, Boccaletti-PhysRep2006,Barrat-PNAS2004,Bianconi-Plone2014}.\\
As well known, the topology and the distribution of weights have strong influence on the dynamical properties of the structure and on the dynamical properties of processes embedded in the structure itself. This is the case, for instance, for the dynamics of polymer networks (e.g., their response to external forces \cite{Gurtovenko-AdvPolymSci2005}), for diffusion and transport processes (e.g., epidemics in social networks \cite{Vespignani-PRL2001}, chemical-kinetics \cite{ABCN-TCA2007}, quantum search algorithms \cite{Agliari-PRA2010}), for statistical mechanics models (e.g., ferromagnetic systems \cite{Barrat-2008,Agliari-EPL2011}, glasses \cite{Castellana-PRL2010} and neural networks \cite{ABGGTT-PRL2015}). 

Now, the topology and the pattern of weights of a given structure can be mathematically captured in terms of the Laplacian matrix, and most dynamical properties can be fully described through the Laplacian eigenvectors and eigenvalues (see e.g., the reviews \cite{Biggs-1993,Mohar-DM1992,Doyle-1984,Alavi-1991,Chung-1994,Mohar-1997}). Also, recent developments have shown how to approach analytically the solution of spin systems via methods based on the extension of the replica method where both the degree of nodes (via ``hard constraints'') and the matrix spectrum (via a ``soft constraint'') are prescribed \cite{Coolen-2016}.
\newline
Beyond the above mentioned physically-driven applications, the study of the graph spectrum realizes increasingly rich connections with computer science (see e.g., \cite{Simic-LAA2011}) and with many other areas of mathematics such as differential geometry (see e.g., \cite{Chung-1994,Das-CMA2004}). 
As a result,  during the past few decades, the study of Laplacian eigenvalues has attracted an upsurge of interest.
Among the (deterministic) structures for which the full knowledge of the spectrum has been achieved we can mention several examples of fractals \cite{Cosenza-PRA1992,Jayanthi-PRL1992}, of small-world networks \cite{Zhang-SciRep2015}, and of scale-free graphs \cite{Zhang-PRE2013b}, just to cite a few. 
However, in general, deriving the exact Laplacian spectrum for an arbitrary system is a challenging task and the use of deterministic structures is of much help to this aim.

Here we focus on a particular example of weighted graph, referred to as $\mathcal{G}$, and we derive analytically its Laplacian spectrum.
More precisely, the graph considered is a weighted, fully-connected network, originally introduced by Dyson \cite{Dyson-CMP1969} to mimic a one-dimensional structure with long-range interactions. In fact, the weight $J_{ij}$ associated to the link connecting node $i$ and node $j$ scales as $J_{ij} \sim 4^{- d_{ij} \sigma}$, where $d_{ij}$ is the distance between the nodes considered and $\sigma$ is a positive tunable parameter that rules the decay of the interaction between nodes as their distance varies. This network is built deterministically and recursively, and we show that, given its building procedure, the coupling matrix $\mathbf{J}$ exhibits a block form which allows the analytical investigation of its spectrum. In fact, we are able to derive the full, exact Laplacian spectrum as well as its eigenvectors.

These findings can be exploited in a wide range of problems, from structural problems (see e.g., \cite{Mieghen-2011}), such as the determination of the spanning trees, of the resistance distance and of the community structures, to dynamical problems as those mentioned above. 
In particular, here, we address a few applications concerning several different fields to show the effectiveness of the Laplacian spectrum.
\newline
First, we calculate the mean first-passage time for a random walker embedded in $\mathcal{G}$, finding that 
the average mean time to first reach a given node depends functionally on $\sigma$ and it grows as the inhomogeneity of the pattern of weights is enhanced.
\newline
Then, we consider a quantum particle moving in a potential described by $\mathcal{G}$ (namely the Hamiltonian which determines the time evolution is identified by the Laplacian of $\mathcal{G}$) and we calculate the long-time average, finding that, in the limit of large size, it converges to $1/3$, much above the equipartition limit $1/N$ expected for classical propagation, hence suggesting that $\mathcal{G}$ is not well performing as far as coherent propagation is concerned.
\newline
Next, we consider $\mathcal{G}$ in the framework of generalized Gaussian structures and, based on the spectra, we calculate the structural average of the mean monomer displacement under applied constant force and the mechanical relaxation moduli, again highlighting the role of inhomogeneity in the coupling pattern. 
\newline
Finally, we exploit our results on the spectra of $\mathcal{G}$ to investigate its community structure. In fact, since the graph under study is regular, the knowledge of the second largest Laplacian eigenvalue allows us to estimate its modularity as a function of $\sigma$ and of the partition considered. 

The paper is organized as follows. 
In the section ``The Laplacian spectrum and its applications'' we review the basic definitions concerning the Laplacian matrix and in the section ``Description of the hierarchical network'' we review the hierarchical structure considered, discussing its building procedure and the block structure of its coupling matrix. Then, in the sections ``Eigenvalues of the Dyson hierarchical graph'' and ``Eigenvectors'' we derive analytically the eigenvalues and the eigenvectors, respectively, of the Laplacian matrix for the graph considered, while in the section ``Examples of applications'' we exploit these findings to derive some results concerning applications in several fields. Finally, the section ``Conclusions'' is left for conclusions, remarks and outlooks.

\section*{The Laplacian spectrum and its applications \label{sec:theory}} 
The Laplacian matrix has a long history in science and there are several books and survey papers dealing with its mathematical properties and applications, see e.g., \cite{Biggs-1993,Mohar-DM1992,Doyle-1984,Alavi-1991,Chung-1994,Mohar-1997}.
Here we just review some basic definitions in order to provide a background for the following analysis.

Let $\textbf{J}$ be the generalized adjacency matrix (weight matrix) of an arbitrary graph $\mathcal{H}$ of size $N$, in such a way that the entry $J_{ij}$ of 
$\textbf{J}$ is non null if $i$ and $j$ are adjacent in $\mathcal{H}$, otherwise $J_{ij}$ is zero. We call $w_i$ the weighted degree of node $i$ which is defined as $w_i \equiv \sum_{j=1}^N J_{ij}$.
Further, let $\mathbf{W}$ be the diagonal matrix, whose entries are given by $W_{ij} = w_i \delta_{ij}$.
The Laplacian matrix of $\mathcal{H}$ is then defined as 
\be \label{eq:Laplacian}
\mathbf{L} = \mathbf{W} - \mathbf{J}. 
\ee

Given that $\mathbf{L}$ is semi-definite positive, the Laplacian eigenvalues are all real and non-negative. Also, they are contained in the interval $[0, \min \{ N,  2 w_{\max} \} ]$, where $w_{\max} \equiv \max_i \{ w_i\}$. The set of all $N$ Laplacian eigenvalues $\varphi_1 \leq \varphi_{2} \leq ... \leq \varphi_N$ is called the Laplacian spectrum. 
Following the definition (\ref{eq:Laplacian}), the smallest eigenvalue $\varphi_1$ is always null and corresponds to the eigenvector $\mu_1 = \mathbf{e}_N$, according to the Perron-Frobenius theorem.
The second smallest eigenvalue is $\varphi_{2} \geq 0$, and it equals zero only if the graph is disconnected. Thus, the multiplicity of $0$ as an eigenvalue of $\mathbf{L}$ corresponds to the number of components of $\mathcal{H}$.

The smallest non-zero Laplacian eigenvalue is often referred to as ``spectral gap'' (or also ``algebraic connectivity'') and it provides information on the effective bipartitioning of a graph.
Basically, a graph with a ``small'' first non-trivial Laplacian eigenvalue has a relatively clean bisection (i.e., the smaller the spectral gap and the smaller the relative number of edges required
to be cut away to generate a bipartition), conversely, a large spectral gap characterizes
non-structured networks, with poor modular structure (see e.g., \cite{Girvan-PNAS2002,Donetti-JStat2004}).\\
The spectral gap has also remarkable effects on dynamical processes. For instance, let us consider a system of $N$ oscillators represented by the nodes of $\mathcal{H}$, which are interconnected pairwise by means of active links; the time evolution of the $i$-th oscillator is given by $\dot{x}_i =  F(x_i) - \beta \sum_{j=1}^N L_{ij} H(x_j),$
where $F$ and $H$ are the evolution and the coupling functions, respectively, and $\beta$ is a coupling constant.
In this case a network exhibits good synchronizability if the eigenratio $\varphi_N/\varphi_2$ is as small as possible and a small spectral gap is therefore likely to imply a poor synchronizability \cite{Wang-IJBC2002}.
\newline
However, probably the easiest dynamical process affected by the underlying topology is the random walk.
In this context, the spectral gap is associated
with spreading efficiency: random walks move around quickly and disseminate fluently
on graphs with large spectral gap in the
sense that they are very unlikely to stay long within a given subset of vertices unless
its complementary subgraph is very small \cite{Lovasz-1993}.

Finally, we mention at the relation between the number of spanning trees in a graph $\mathcal{H}$ and its Laplacian spectrum.
Let us denote with 
$\Omega(\mathcal{H})$ the number of spanning trees in $\mathcal{H}$:
the Kirchhoff matrix-tree theorem states that (see e.g., \cite{Harris-2008,Chang-PJM2015})
$\Omega(\mathcal{H}) = N^{-1} \prod_{k=2}^{N} \varphi_k$.
This formula turns out to be extremely useful in order to get bounds for $\Omega(\mathcal{H})$ even without having the exact spectrum of the related graph, or in order to estimate the number of spanning trees of complex graphs which can be defined as combinations of simpler graphs for which the exact spectrum is known (see e.g., \cite{Li-AML2010}).\\
For a weighted graph where $w_{ij} \in \mathbb{N}$ is the number of edges joining $i$ and $j$, the previous formula for $\Omega(\mathcal{H})$ still holds. More generally, when $w_{ij} \in \mathbb{R}$ one can still look for the number of spanning trees on the bare topology (i.e., neglecting weights), or look for the minimum spanning tree(s), where the sum of the weights over all the edges making up the tree(s) is minimal (see e..g., \cite{Harary-1969})\\
Beyond the purely mathematical point of view, deriving $\Omega(\mathcal{H})$ is a key problem in many areas of experimental design for the synthesis of reliable communication networks where the links of the network are subject to failure \cite{Bondy-1976}. 
Also, the number of spanning trees is related to the configurational integral (or partition function) $Z$ of a Gaussian macromolecule, whose architecture is described by $\mathcal{H}$; in fact, one has $Z \propto [\Omega(\mathcal{H}) ]^{-3/2}$, where the proportionality constant does not depend on the topology but just on the number of monomers, on the temperature and on the spring constant between beads \cite{Eichinger-Macro1980}.\\

Further applications for the Laplacian spectrum will be reviewed and deepened in the section ``Examples of applications'' focusing on the hierarchical Dyson network.

\section*{Description of the hierarchical network \label{sec:Hierarchical} }

In this work we focus on a deterministic, weighted, recursively grown graph, referred to as $\mathcal{G}$, originally introduced by Dyson to study the statistical-mechanics of spin systems beyond the mean-field scenario (corresponding to a fully-connected, unweighted embedding) \cite{Dyson-CMP1969}. The topological properties of this graph have been discussed in \cite{ABGGTT-PRL2015,ABGGTT-PRE2015,TA-PRE2016}, and here we briefly review them. 
The construction begins with $2$ nodes,  connected with a link carrying a weight $J(1,1,\sigma) = 4^{-\sigma}$. We refer to this graph as $\mathcal{G}_1$, and in the notation $J(d,k,\sigma)$ we highlight the dependence on the graph iteration $k$ and on the system parameter $\sigma$, also, $d$ represents the iteration when the nodes considered first turn out to be connected.  At the next step, one takes two replicas of $\mathcal{G}_1$ and connects the nodes pertaining to different replicas with links displaying a weight $J(2,2,\sigma) = 4^{-2 \sigma}$; moreover, the weight on the existing links is updated as $J(1,1,\sigma) \rightarrow J(1,2,\sigma) = J(1,1,\sigma) + J(2,2,\sigma)$. This realizes the graph $\mathcal{G}_2$, which counts overall $4$ nodes.  At the generic $k$-th iteration, one takes two replicas of $\mathcal{G}_{k-1}$, insert $2^{2k-1}$ new links, each carrying a weight $J(k,k,\sigma)=4^{-k \sigma}$, among nodes pertaining to different replicas, and the weights on existing links are updated as $J(d,k-1,\sigma) \rightarrow J(d,k,\sigma) = J(d,k-1,\sigma) + J(k,k,\sigma)$, for any $d<k$. If we stop the iterative procedure at the $K$-th iteration, the final graph $\mathcal{G}_{K}$ counts $N=2^K$ nodes and the coupling between any pair of nodes can be expressed as
\begin{equation}\label{HPScoupling}
J(d,K,\sigma)=\sum_{l=d}^{K}J(l,K,\sigma)=\sum_{l=d}^{K}4^{-l\sigma}=\frac{4^{\sigma(1-d)}-4^{-K\sigma}}{4^{\sigma}-1},
\end{equation}
with $d \in \{ 1, ..., K\}$.
Remarkably, this iterative procedure allows for a definition of metric:  two nodes are said to be at distance $d$ if they occur to be first connected at the $d$-th iteration [Note: One can check that this metric is intrinsically ultrametric since, beyond the standard conditions for a well defined metric ($d_{ij} \geq 0$; $d_{ij}=0 \Leftrightarrow i=j$; $d_{ij}=d_{ji}$), the so-called ultrametric
inequality  ($d_{ij} \leq \max (d_{iz},d_{zj})$) also holds.]. As a result, we can define a coupling matrix $\mathbf{J}$ associated to the network $\mathcal{G}_K$ such that the entry $J_{ij}$ depends on the nodes $(i,j)$ considered only through their distance $d_{ij}$, namely  
\begin{equation} \label{eq:coupling}
 J_{ij}=\frac{4^{\sigma(1-d_{ij})}-4^{-K\sigma}}{4^{\sigma}-1}.
\end{equation} 
Given the building procedure of $\mathcal{G}_K$, a generic node $i$ has $2^{d-1}$ nodes at distance $d\in \{1,...,K \}$.
Moreover, the total weight of the links stemming from a single node $i$ can be written as
\begin{eqnarray}
w_{i}&= &\sum_{i\neq j}J_{ij}=\sum_{l=1}^{K}2^{d-1}J(d,K,\sigma)=\nonumber\\
&=&\frac{2N^{1-2\sigma}(1-2^{2\sigma})+N^{-2\sigma}(2^{2\sigma}-2)+2^{2\sigma}}{(2^{2\sigma}-1)(2^{2\sigma}-2)}.\label{weight}
\end{eqnarray}
Due to the symmetry underlying the network, $w_i$ does not depend on the site $i$, so one can simply write $w_i=w$.

Beyond the coupling matrix $\mathbf{J}$, one can introduce the Laplacian matrix $\mathbf{L}$, which, as anticipated in Eq.~\ref{eq:Laplacian}, is defined as $\mathbf{L} = \mathbf{W} - \mathbf{J}$, where $\mathbf{W}$ is a diagonal matrix with elements $W_{ij} = w \delta_{ij}$. 

{Before concluding this section it is worth stressing that the parameter $\sigma$ is bounded as $1/2 < \sigma \leq 1$. In a statistical mechanics context this ensures that the Dyson model is thermodynamically well defined \cite{Castellana-PRL2010,ABGGTT-PRL2015}; in this context the lower bound $\sigma > 1/2$ ensures that the weight $w$ remains finite in the limit $N \rightarrow \infty$, while the upper bound $\sigma \leq 1$ ensures that the heterogeneity in the coupling pattern is not too strong (namely, that the relaxation time, given by the inverse of the spectral gap, does not grow faster than the system size).}

\section*{Eigenvalues of the Dyson hierarchical graph}\label{sec:eig1}

In order to get familiar with the structure of  the matrix $\mathbf{J}$, it is convenient to write it down explicitly for a small value of $K$. In particular, for $K=3$ it reads as 
\be
\mathbf{J}^{(N)}=\left( \begin{array}{cccc|cccc}
	0 & \sum_{i=1}^{3}t^i & \sum_{i=2}^{3}t^i & \sum_{i=2}^{3}t^i & t^3 & t^3 & t^3 & t^3\\
	\sum_{i=1}^{3}t^i & 0 & \sum_{i=2}^{3}t^i & \sum_{i=2}^{3}t^i & t^3 & t^3 & t^3 & t^3\\
	\sum_{i=2}^{3}t^i & \sum_{i=2}^{3}t^i & 0 & \sum_{i=1}^{3}t^i & t^3 & t^3 & t^3 & t^3\\
	\sum_{i=2}^{3}t^i & \sum_{i=2}^{3}t^i & \sum_{i=1}^{3}t^i & 0 & t^3 & t^3 & t^3 & t^3\\
	\hline
	t^3 & t^3 & t^3 & t^3 &	0 & \sum_{i=1}^{3}t^i & \sum_{i=2}^{3}t^i & \sum_{i=2}^{3}t^i\\
	t^3 & t^3 & t^3 & t^3 &	 \sum_{i=1}^{3}t^i & 0 & \sum_{i=2}^{3}t^i & \sum_{i=2}^{3}t^i\\
	t^3 & t^3 & t^3 & t^3 & \sum_{i=2}^{3}t^i & \sum_{i=2}^{3}t^i & 0 &  \sum_{i=1}^{3}t^i\\
	t^3 & t^3 & t^3 & t^3 & \sum_{i=2}^{3}t^i & \sum_{i=2}^{3}t^i  &\sum_{i=1}^{3}t^i & 0 \end{array} \right),\label{blockmat}
\ee
where we posed $t=4^{-\sigma}$ and we highlighted with the superscript $(N)$ the size of the matrix; in general, $N=2^K$ and here $N=8$.
The block structure is evident and can be schematised as
\be
\mathbf{J}^{(N)}=\left( \begin{array}{c|c}
	\mathbf{L}{^{(N/2)}} & \mathbf{R}^{(N/2)}\\
	\hline
	\mathbf{R}{^{(N/2)}} & \mathbf{L}{^{(N/2)}}
\end{array} \right),
\ee
where  $\mathbf{L}^{(N/2)}$ and  $\mathbf{R}^{(N/2)}$ are square matrices of size $\frac{N}{2} \times\frac{N}{2}$. For such matrices, the determinant can be computed as
\be \label{eq:Jrec}
\det(\mathbf{J}^{(N)})=\det(\mathbf{L}{^{(N/2)}}-\mathbf{R}{^{(N/2)}})\det(\mathbf{L}{^{(N/2)}}+\mathbf{R}{^{(N/2)}}).
\ee
Using the ultrametric structure of $\mathbf{J^{(N)}}$, we can iterate this block decomposition. In fact, for example, after the first decomposition we obtain two matrices $\mathbf{J}_1^{N/2}$ and $\mathbf{J}_2^{N/2}$ such that
\be
\mathbf{J_1}{^{(N/2)}} \equiv \mathbf{L}{^{(N/2)}} - \mathbf{R}{^{(N/2)}}=\left( \begin{array}{c|c}
	\mathbf{L_1}{^{(N/4)}} & \mathbf{R_1}{^{(N/4)}}\\
	\hline
	\mathbf{R_1}{^{(N/4)}} & \mathbf{L_1}{^{(N/4)}}
\end{array} \right),
\qquad
\mathbf{J_2}{^{(N/2)}} \equiv \mathbf{L}{^{(N/2)}} + \mathbf{R}{^{(N/2)}}=\left( \begin{array}{c|c}
	\mathbf{L_2}{^{(N/4)}} & \mathbf{R_2}{^{(N/4)}}\\
	\hline
	\mathbf{R_2}{^{(N/4)}} & \mathbf{L_2}{^{(N/4)}}
\end{array} \right).
\ee
Now, both $\mathbf{J_1}{^{(N/2)}}$ and $\mathbf{J_2}{^{(N/2)}}$ display the same block structure and for each an expression analogous to (\ref{eq:Jrec}) holds in such a way that 
\be
\det(\mathbf{J}^{(N)})=\det(\mathbf{L_1}^{(N/4)}-\mathbf{R_1}^{(N/4)})\det(\mathbf{L_1}^{(N/4)}+\mathbf{R_1}^{(N/4)})\det(\mathbf{L_2}^{(N/4)}-\mathbf{R_2}^{(N/4)})\det(\mathbf{L_2}^{(N/4)}+\mathbf{R_2}^{(N/4)}).
\ee
At this point, we pose
\begin{eqnarray}
	\mathbf{J_1}^{(N/4)} &\equiv&	\mathbf{L_1}^{(N/4)}-\mathbf{R_1}^{(N/4)}\nonumber\\
	\mathbf{J_2}^{(N/4)} &\equiv&	\mathbf{L_1}^{(N/4)}+\mathbf{R_1}^{(N/4)}\nonumber\\
	\mathbf{J_3}^{(N/4)} &\equiv&	\mathbf{L_2}^{(N/4)}-\mathbf{R_2}^{(N/4)}\nonumber\\
	\mathbf{J_4}^{(N/4)} &\equiv&	\mathbf{L_2}^{(N/4)}+\mathbf{R_2}^{(N/4)}\nonumber
\end{eqnarray} 
where, as stated previously, the superscript indicates the size of the matrices and, proceeding iteratively, we get a set of $N/2$ matrices of size $2\times 2$ and referred to as $\{\mathbf{J}^{(2)}_l\}$ with $l=1,...,N/2$.
\newline
More precisely, at every $n$-th iteration, we can write $2^n$ block matrices $\{\mathbf{J}^{(N/2^n)}_l\}_{l=1}^{2^n}$ of size $N/2^{n}$, $n=1,...K-1$, obtained as the sum or the difference of the $2^{n-1}$ matrices of the  previous iteration. With this argument, we can state that our determinant is a product of $N/2$ determinants of matrices $2\times 2$ as
\be
\det(\mathbf{J}^{(N)})=\prod_{n=1}^{N/2}\det{\mathbf{J}_n}^{(2)}.
\ee
Clearly, in order to compute the eigenvalues of $\mathbf{J}$, we can again use this scheme, but taking as initial matrix $\mathbf{J}-\lambda\mathbf{I}$. In this case, with some algebra, we obtain the following
\be \label{eq:equazioneautovalori}
\det(\mathbf{J}-\lambda\mathbf{I})=\prod_{n=1}^{K}p_n,
\ee
where $p_n$ is the determinant of a proper matrix of size $N/2^n \times N/2^n$ and it reads as
\be \label{eq:equazioneautovalori2}
p_n=\left\{ \left[ -\lambda+\sum_{l=2}^{n}(2^{l-1}-1)t^l-\sum_{l=n+1}^{K}t^l \right]^2- \left(\sum_{l=1}^{n}2^{l-1}t^l \right)^2\right\}^{\frac{N}{2^{n+1}} }, \qquad n=1,\cdots,K.
\ee
By plugging the previous expression for $p_n$ into Eq.~\ref{eq:equazioneautovalori} and solving for the roots of $\det(\mathbf{J}-\lambda\mathbf{I})=0$, we get $K+1$ distinct eigenvalues that can be written as
\begin{eqnarray}
\label{eq:star}
\lambda_n&=&\frac{N^{-2\sigma}(2^{2\sigma}-2)-2\times 2^{n(1-2\sigma)}(2^{2\sigma}-1)+2^{2\sigma}}{(2^{2\sigma}-1)(2^{2\sigma}-2)}, ~~ n=0,\cdots,K-1, \text{ with algebraic multiplicity }\frac{N}{2^{n+1}},\\
\label{eq:star2}
\lambda_K&=& w,\text{ with algebraic multiplicity $1$} ,
\end{eqnarray}
Notice that, as $n$ increases, both $\lambda_n$ and its multiplicity decreases.

\begin{figure}[tb]
	\includegraphics[width=12cm]{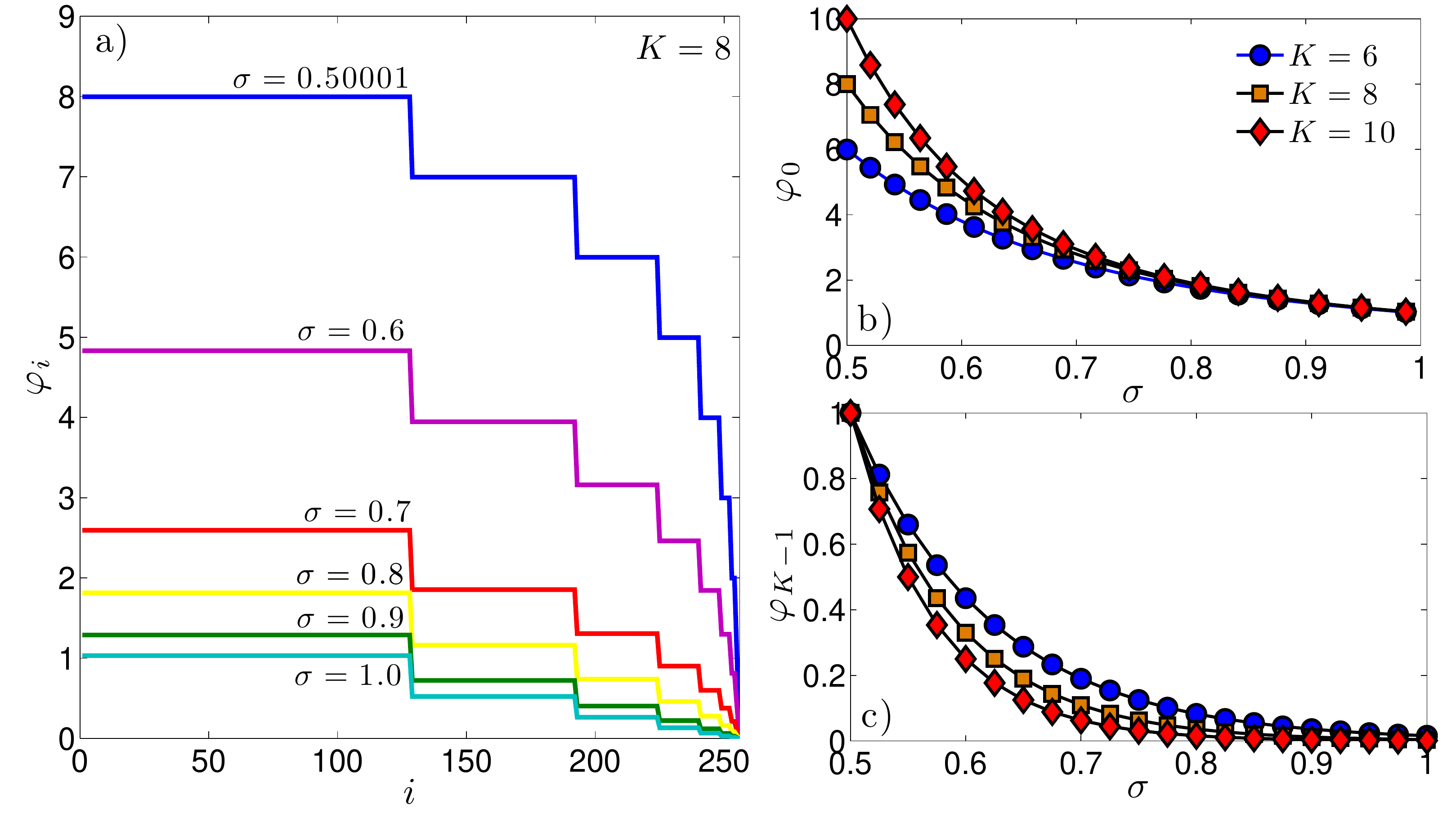}
	\caption{Panel $a$: Eigenvalue spectrum of $\mathbf{L}$ for $K=8$ and for different values of $\sigma$. The multiplicity of an eigenvalue corresponds to the width of the related plateau and it grows with the magnitude of the eigenvalue. Moreover, the spectrum is broadened over a range which decreases with $\sigma$. Panel $b$: trend of the eigenvalue $\varphi_0$ when $\sigma$ varies in the interval $(1/2,1]$ for different values of $K$ as shown in the legend. Since $\varphi_0$ corresponds to the largest eigenvalue this also provides the span of the spectrum itself (notice that the smallest eigenvalue $\varphi_K$ is always null). Panel $c$: trend of the eigenvalue $\varphi_{K-1}$ when $\sigma$ varies in the interval $(1/2,1]$ for different values of $K$ (the legend is the same as in panel $b$). Since $\varphi_{K-1}$ corresponds to the smallest non-null eigenvalue this also provides the spectral gap for $\mathcal{G}_K$. }\label{trendE}
\end{figure}

Actually, we are mainly interested in the spectrum of the Laplacian matrix $\textbf{L}=\mathbf{W}-\mathbf{J}$, with $\mathbf{W}=w\delta_{ij}$, where we recall that $w$ is the weighted degree defined in Eq. $\ref{weight}$. Now, denoting with $\varphi_n$ the $n$-th eigenvalue of $\mathbf{L}$, and exploiting the fact that $\mathbf{W}$ is proportional to the identity matrix, one can write
\be
\varphi_n=w-\lambda_n,\qquad n=0,...,K.
\ee
More precisely we have
\begin{eqnarray}
	\varphi_n&=&\frac{2^{n(1-2\sigma)}-N^{1-2\sigma}}{2^{2\sigma-1}-1} \qquad n=0,...,K-1\text{ with algebraic multiplicity }\frac{N}{2^{n+1}},\label{ee} \\
	\varphi_K &=& 0, \text{ with algebraic multiplicity 1}.\label{e}
\end{eqnarray}
Of course, since $\mathbf{L}$ is semidefinite positive, $\varphi_n \geq 0$ $\forall n\geq 0$, and, in particular, omitting the trivial eigenvalue $\varphi_K =0$, we have
\begin{eqnarray}
\label{eq:max}
	\max_{n=0,...,K-1}\varphi_n&=&\varphi_0=(1-N^{1-2\sigma})/(2^{2\sigma-1}-1)\\ 
	\min_{n=0,...,K-1}\varphi_{n}&=&\varphi_{K-1}=N^{1-2\sigma}.
\end{eqnarray}
From Eq.~\ref{ee} we notice that the larger the value of $\varphi_n$, and the larger its algebraic multiplicity (see also Fig.~\ref{trendE}, panel $a$). Moreover, from Eq.~\ref{eq:max} we see that $\varphi_0$ decreases with $\sigma$ (see Fig.~\ref{trendE}, panel $b$) and, as a consequence, {the spectrum covers a narrower interval (we recall that $\varphi_{K}=0$ hence the spectrum width is given by $\varphi_{0}$). However, the eigenratio $\varphi_{0}/\varphi_{K-1}$ (which provides a more effective measure of the spectrum width) grows with $\sigma$ implying poor synchronizability for large $\sigma$. This is consistent with the fact that large values of $\sigma$ correspond to heterogeneous patterns of couplings \cite{TA-PRE2016}. Similarly, the spectral gap, corresponding to $\varphi_{K-1}$, decreases fast with $\sigma$ (see Fig.~\ref{trendE}, panel $c$) and, as a consequence, the spreading efficiency on $\mathcal{G}_K$ is weakened as the patter of couplings gets more heterogeneous}.

\begin{figure}[tb]
	\includegraphics[width=12cm]{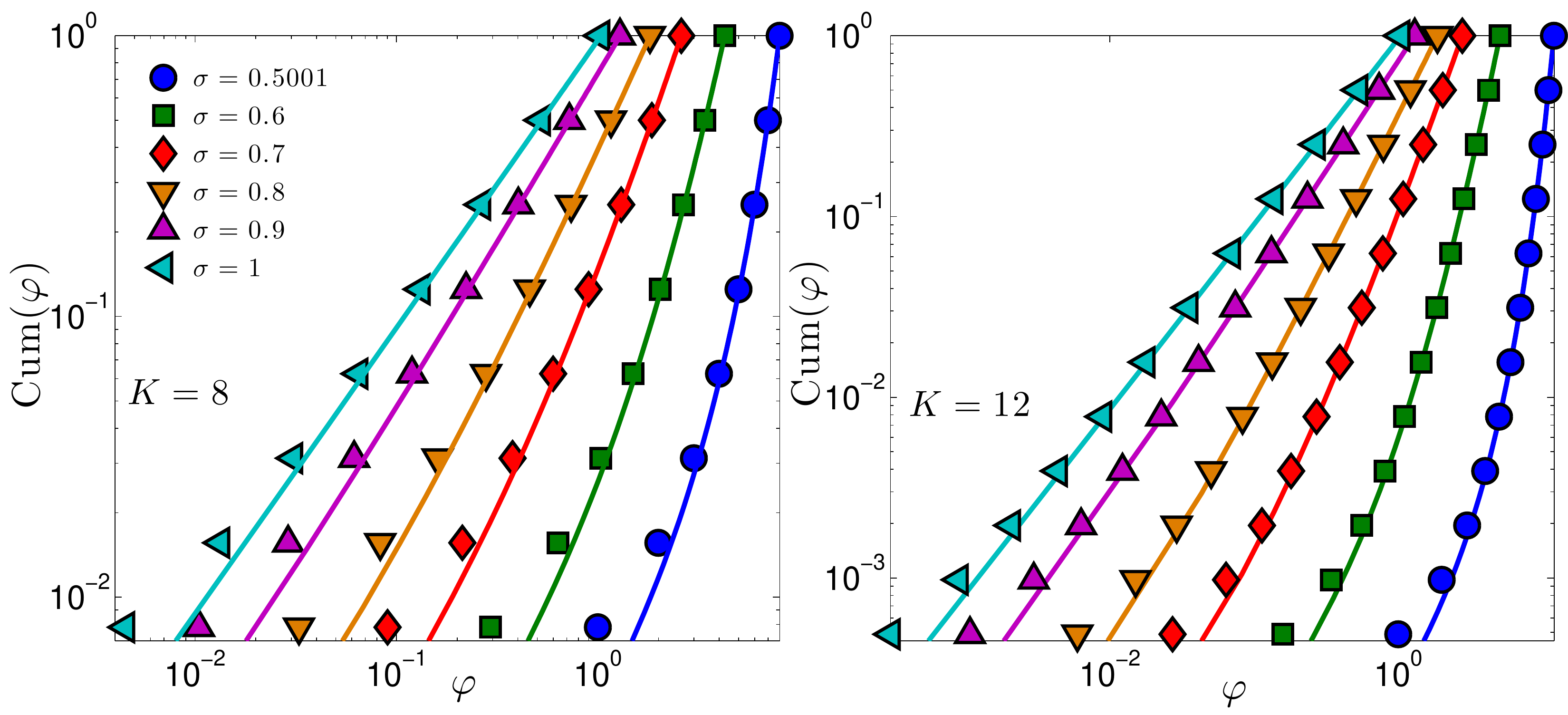}
	\caption{The cumulative function $\text{Cum}(\varphi)$ defined in Eq.~\ref{eq:cum} is shown for $K=8$ (left panel) and $K=12$ (right panel) and for several choices of $\sigma$. Symbols correspond to the numerical estimate of Eq.~\ref{eq:cum}, while solid lines correspond to the theoretical expression in Eq.~\ref{eq:cumul}.}\label{cum}
\end{figure}

Equations \ref{ee} and \ref{e} provide an explicit expression for the spectrum of the Laplacian matrix associated to the hierarchical graph under study.
{The Laplacian spectral density $\rho(\varphi)$, where $\rho(\varphi) d \varphi$ returns the fraction of eigenvalues} laying in the interval $(\varphi , \varphi + d \varphi)$, can be analytically recovered as follows. First, we determine the cumulative distribution $\text{Cum}(\varphi)$: fixing an arbitrary value $\bar{\varphi} \in [\varphi_{K}, \varphi_0 ]$, the overall number of eigenvalues smaller than or equal to $\bar{\varphi}$ is given by
\be \label{eq:cum}
\text{Cum}(\bar{\varphi}) = \frac{1}{N} \sum_{n=\lfloor n(\bar{\varphi}) \rfloor }^{K-1} \frac{N}{2^{n+1}},
\ee
where $n(\bar{\varphi})$ is obtained by inverting Eq.~\ref{ee}, that is,
\be \label{eq:indice}
n(\bar{\varphi})=\frac{1}{1-2\sigma}\log_2\Big[N^{1-2\sigma}+\bar{\varphi}(2^{2\sigma-1}-1)\Big], 
\ee
and where $\lfloor x \rfloor$ denotes the largest integer not greater than $x$. 
We stress that $\lfloor n(\bar{\varphi}) \rfloor$ provides the index $n$ such that $\varphi_n$ is the eigenvalue closest to $\bar{\varphi}$, with $\varphi \leq \bar{\varphi}$. Moreover, we remark that  in Eq.~\ref{eq:cum}, the index of the sum varies between  $\lfloor n(\bar{\varphi}) \rfloor$ and $K-1$: this is because, according to the notation introduced in Eq.~\ref{ee}, the eigenvalues decrease as $n$ increases, so to find all the eigenvalues smaller than or equal to a fixed $\bar{\varphi}$, it is necessary to consider the larger indices.
\newline
By merging the results (\ref{eq:cum}) and (\ref{eq:indice}), we can get a close-form expression for the cumulative distribution as
\be \label{eq:cumul}
\text{Cum}(\varphi)  \approx  \left[N^{1-2\sigma}+\varphi(2^{2\sigma-1}-1) \right]^{\frac{1}{2\sigma-1}}-\frac{1}{N},\qquad\text{with } \varphi_{K-1} \leq \varphi \leq \varphi_{0}.
\ee
Of course, in the limit $N \rightarrow \infty$, $\text{Cum}(\varphi_{0})$ and $\text{Cum}(\varphi_{K-1})$ converge to the expected values, being $\text{Cum}(\varphi_{0})  = 1 -1/N \rightarrow 1$, and $\text{Cum}(\varphi_{K-1}) = 1/N \rightarrow 0$. The expression (\ref{eq:cumul}) is numerically checked in Fig.~\ref{cum}.

\begin{figure}[tb]
	\includegraphics[width=12cm]{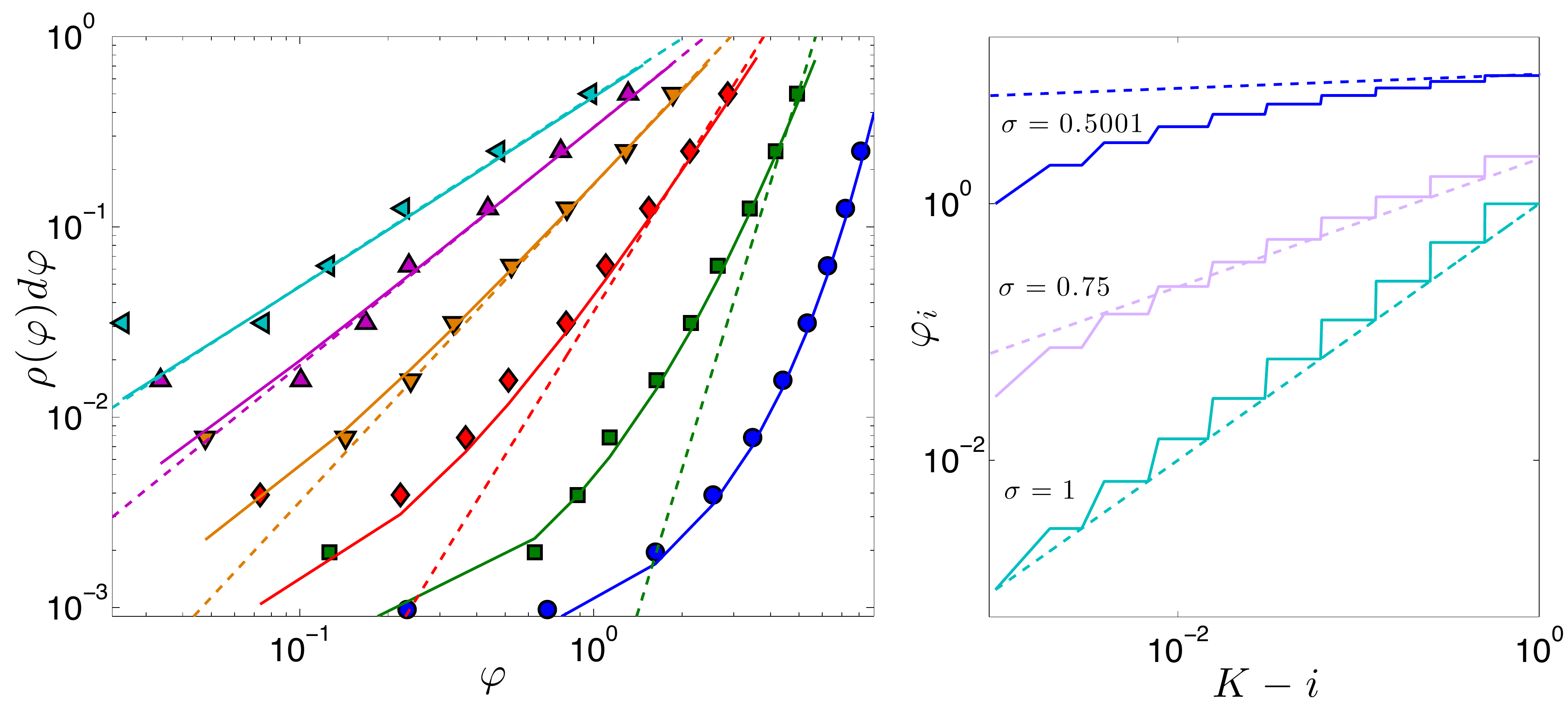}
	\caption{Left panel: The eigenvalue distribution $\rho(\varphi) d \varphi$ for a system of size $N=2^K$ with $K=10$ is shown versus $\varphi$ and for several choices of $\sigma$ (the legend is the same as in Fig.~\ref{cum}). In particular, in this figure we show the histogram (symbols) derived from the exact spectrum, the theoretical estimate (solid line) given by Eq.~\ref{eq:rho}, and the related scaling $\rho(\varphi) d \varphi \sim \varphi^{1/(2\sigma -1)}$ (dashed line). The comparison is successful especially for large values of $\varphi$ and for $\sigma$ not too small. {Right panel: The eigenvalues for a system of size $N=2^K$ with $K=10$ are plotted in ascending order, for several choices of $\sigma$, as reported. The dashed lines highlight the expected behaviour according to the analytical estimate $\sim (K-i)^{2 \sigma -1}$, following from Eq.~\ref{eq:cumul}. Again, one can see that the theoretical picture is quantitatively good just for large values of $\sigma$, while it turns out to be qualitatively correct in the whole range considered. In fact,  a slower rate of growth for $\varphi_i$ versus $K-i$ implies a faster rate of growth for $\rho(\varphi)$ versus $\varphi$, which in turns implies a larger spectral dimension, in agreement with Eq.~\ref{eq:ds} }. }\label{Pphi}
\end{figure}

Finally, taking the derivative of $\text{Cum}(\varphi)$, we get an estimate for $\rho(\varphi)$:
\be \label{eq:cumul2}
\rho(\varphi) d \varphi \approx   - \frac{\partial \text{Cum}(\varphi_n)}{\partial n} dn,
\ee
{and, using  (\ref{eq:cum}) and (\ref{eq:indice}), we get}
\be \label{eq:rho}
\rho(\varphi) \sim 
[ N^{1-2\sigma} + \varphi (2^{2\sigma-1}-1) ]^{\frac{1}{2 \sigma-1}-1}. 
\ee
{Thus, when $\varphi \gg N^{1-2\sigma}$ (i.e., for large sizes, and/or for $\s$ close to $1$, and/or in the upper part of the spectrum)}, the distribution scales as $\rho(\varphi) \sim \varphi^{1/(2\sigma -1)-1}$. This result is checked numerically in Fig.~\ref{Pphi} (left panel) {: as expected, the comparison between the theoretical and the numerical estimates is successful especially in the case $\sigma=1$, while for lower values deviations with respect to the power-law behavior emerge. Interestingly, as long as this picture holds, the expression in Eq.~\ref{eq:rho}
allows us to get an estimate for the spectral dimension $d_s$ of $\mathcal{G}$. In fact, recalling that in the small eigenvalue limit $\varphi \rightarrow 0$ one has $\rho(\varphi) \sim  \varphi^{d_s/2 -1}$, we get}
\be \label{eq:ds}
d_s(\sigma) \approx \frac{2}{2\sigma-1}.
\ee
In particular, $d_s(\sigma=1) \approx 2$, and it monotonically grows as $\sigma$ approaches its lower bound. In fact, as $\sigma$ is reduced, the pattern of couplings gets more and more homogenous towards a mean-field scenario \cite{Castellana-PRL2010, TA-PRE2016}. 
This result is further checked in Fig.~\ref{Pphi} (right panel). We stress that the spectral dimension (when defined) generalises the Euclidean dimension to the case of non-translationally-invariant structures: in fact, this is a global property of the graph, related (beyond to the density of small eigenvalues of the Laplacian), for instance, to the infrared singularity of the Gaussian process, or to the (graph average) of the long-time tail of the random walk return probability; it also provides a consistent criterion on whether a continuous symmetry breaks down at low temperature (see e.g., \cite{Agliari-2016}).

Finally, it is worth comparing the results found here for $\mathcal{G}$ with those pertaining to other examples of graphs. For instance, if we neglect the weights in $\mathcal{G}$ we recover the complete graph (cg) of size $N$, often referred to as $K_N$, where the Laplacian spectrum is given by $\varphi^{\mathrm{(cg)}} = 0$ with multiplicity $1$ and $\varphi^{\mathrm{(cg)}} = N$ with multiplicity $N-1$. Beyond this peculiar case, as anticipated in the Introduction, there are several other examples of deterministically grown networks, exhibiting a non-trivial topology, for which the exact spectrum is known. In particular, we mention the Vicsek fractals (vf), where the spectral dimension is given by $d_s^{\mathrm{(vf)}} = 2 \log (f+1) / \log(3f +3) \in [0,2)$, with $f$ being a parameter which sets the coordination number of inner points \cite{Blumen-Macro2004}, and a class of small-world networks (sw) obtained by recursively joining together $K_d$ graphs, where the spectral dimension is given by $d_s^{\mathrm{(sw)}} = 2$, regardless of $d$ \cite{Zhang-SciRep2015}. For relatively sparse graphs, such as exactly decimable fractals (e.g., Sieprinski gasket, T-fractal), one can prove that $d_s < 2$ \cite{Redner-2001Book,Agliari-2016}.

\section*{Eigenvectors \label{sec:eig2}}

Having computed the spectrum of $\mathbf{L}$, we can now proceed with the calculation of its eigenvectors. As shown in the previous section, we have $K+1$ distinct eigenvalues, each with its own algebraic multiplicity. We are looking for a system of linear independent vectors $\{\mu_i\} _{i=1}^N$, where $\mu_i$ is a $N\times 1$ vector such that
\begin{equation}
	\mathbf{L}\mu_n=\varphi_n\mu_n\qquad n=1,..,N,\label{eigenvaluesformula}
\end{equation}
with $\mu_i\mu_j=\delta_{ij}$ and $ \parallel \mu_i\parallel=1$, $\forall i=1,...,N$.

Again, we can exploit the block structure of $\mathbf{L}$ to find these eigenvectors. In fact, for $K=3$, and posing again $t=4^{-\sigma}$, $\mathbf{L}$ has the form
\be
\mathbf{L}=\left( \begin{array}{cccccccc}
	w & -\sum_{i=1}^{3}t^i & -\sum_{i=2}^{3}t^i & -\sum_{i=2}^{3}t^i & -t^3 & -t^3 & -t^3 & -t^3\\
	-\sum_{i=1}^{3}t^i & w & -\sum_{i=2}^{3}t^i & -\sum_{i=2}^{3}t^i & -t^3 & -t^3 & -t^3 & -t^3\\
	-\sum_{i=2}^{3}t^i & -\sum_{i=2}^{3}t^i & w & -\sum_{i=1}^{3}t^i & -t^3 & -t^3 & -t^3 & -t^3\\
	-\sum_{i=2}^{3}t^i & -\sum_{i=2}^{3}t^i & -\sum_{i=1}^{3}t^i & w & -t^3 & -t^3 & -t^3 & -t^3\\
	-t^3 & -t^3 & -t^3 & -t^3 &	w & -\sum_{i=1}^{3}t^i & -\sum_{i=2}^{3}t^i & -\sum_{i=2}^{3}t^i\\
	-t^3 & -t^3 & -t^3 & -t^3 &	 -\sum_{i=1}^{3}t^i & w & -\sum_{i=2}^{3}t^i & -\sum_{i=2}^{3}t^i\\
	-t^3 & -t^3 & -t^3 & -t^3 & -\sum_{i=2}^{3}t^i & -\sum_{i=2}^{3}t^i & w &  -\sum_{i=1}^{3}t^i\\
	-t^3 & -t^3 & -t^3 & -t^3 & -\sum_{i=2}^{3}t^i & -\sum_{i=2}^{3}t^i  &-\sum_{i=1}^{3}t^i & w \end{array} \right).
\ee

Let us start with the eigenvalue $\varphi_{0} = (1-N^{1-2\sigma})/(2^{2\sigma-1}-1)$ (see Eq.~\ref{ee}): it has algebraic multiplicity $N/2$, so we have to find $N/2$ linear independent eigenvectors $\{\mu_j\}_{j=1}^{\frac{N}{2}}$ of size $N\times 1$, associated to this eigenvalue such that
\begin{equation}
	\mathbf{L}\mu_{j}=\left(\frac{1-N^{1-2\sigma}}{2^{2\sigma-1}-1}\right)\mu_{j},\qquad \forall j=1,\cdots, \frac{N}{2}.\label{ansatz1}
\end{equation}
Our ansatz is 
\be \label{eq:ansatz}
\mu_j=(0,\cdots,0,\overbrace{\frac{1}{\sqrt{2}}}^{2j-1},\overbrace{-\frac{1}{\sqrt{2}}}^{2j},0,\cdots,0),\qquad j=1,...,N/2
\ee
that is, a vector that has all the entries equal to zero, but those at the position $2j-1$ an $2j$, that are respectively equal to $+1/\sqrt{2}$ and $-1/\sqrt{2}$, in order to obtain $N/2$ vectors with norm equal to $1$. In this case, the left hand side of Eq.~\ref{eigenvaluesformula} becomes
\begin{equation}
	(\mathbf{L}\mu_j)_i =\begin{cases} \frac{(-1)^i}{\sqrt{2}}(L_{2j-1,2j-1}-L_{2j-1,2j}), & \mbox{if } i=2j-1,2j, \\ 0, & \mbox{otherwise }.
	\end{cases} 
	\label{a}
\end{equation}
Exploiting the structure of $\mathbf{L}$, for every row $i=1,...,N$, we have
\begin{eqnarray}
	L_{2j-1,2j-1}-L_{2j-1,2j}&=&(-1)^i\Big( w+\sum_{l=1}^{K}4^{-l\sigma}\Big)=\frac{1-N^{1-2\sigma}}{2^{2\sigma-1}-1}, \qquad \forall j=1,...,N/2.\nonumber\label{difference}
\end{eqnarray}

Comparing this expression with  ($\ref{ansatz1}$) we get that the ansatz (\ref{eq:ansatz}) is correct. 
In this way we find a set of $N/2$ independent eigenvectors associated to the eigenvalue $\varphi_{0}$. 
\newline
We are going to proceed analogously for the computation of the eigenvectors related to  $\varphi_n=(2^{n(1-2\sigma)}-N^{1-2\sigma})/(2^{2\sigma-1}-1)$ for $n=1,...,K-1$, each with multiplicity $N/2^{n+1}$.  This implies that each of them is associated to $2^{K-n-1}$ linear independent eigenvectors. In particular, we claim that they have the form 
\begin{equation}
	\mu_j=(0,\cdots,0,\overbrace{\frac{1}{\parallel \textbf{e} \parallel}}^{1+i2^{n+1}},...,
	\overbrace{\frac{1}{\parallel \textbf{e}\parallel}}^{(i+1)2^{n+1}},
	\overbrace{-\frac{1}{\parallel \textbf{e}\parallel}}^{1+(i+1)2^{n+1}},...,
	\overbrace{-\frac{1}{\parallel \textbf{e}\parallel}}^{(i+1)2^{n+2}},
	0,\cdots,0),\qquad i=0,...,n-K-1,\qquad j=1,...,2^{K-n-1},\label{ansatz2}
\end{equation}
where $\mathbf{e}$ is a one vector of size $2^n$ such that $\parallel \textbf{e} \parallel=2^{n/2}$ and its first entry is either at the position $1+i2^n$, or at the position $1+(i+1)2^n$ of $\mu_j$ with $i=0,...,n-K-1$ and $j=1,...,2^{K-n-1}$. With some algebra (as done previously in order to get (\ref{a}) from (\ref{eq:ansatz})), we obtain
\begin{equation}
	(\mathbf{L}\mu_j)_i=\begin{cases} \frac{(-1)^i}{2^{\frac{n}{2}}}\Big(\sum_{l=1+i2^n}^{(i+1)2^n}L_{j,l}-\sum_{l=1+2^{n}(i+1)}^{2^{n+1}(i+1)}L_{j,l}\Big), & \mbox{if } j=1+i2^{n+1},...,(i+1)2^{n+2}, \\\\ 0 & \mbox{otherwise }.
	\end{cases}\label{eqEig}
\end{equation}
where $ n=1,..,K-1$, and $ i=0,...,K-n-1$.
The previous equation can be further simplified if one recasts the right-hand term of $(\ref{eqEig})$ as
\begin{eqnarray}
	\sum_{l=1+i2^n}^{(i+1)2^n}L_{j,l}-\sum_{l=1+2^{n}(i+1)}^{2^{n+1}(i+1)}L_{j,l}&=&(-1)^i\Big( w-\sum_{l=1}^{n}2^{l-1}4^{-l\sigma}+\sum_{l=n+1}^{K}2^{l-1}4^{-l\sigma}\Big)\nonumber\\
 	&=&(-1)^i\Big(\sum_{l=n+1}^{K}2^{l(1-2\sigma)}\Big)\nonumber\\
	&=&(-1)^i\Big(\frac{2^{n(1-2\sigma)}-N^{1-2\sigma}}{2^{2\sigma-1}-1}\Big).\nonumber
\end{eqnarray}
The last expression can be compared to $\varphi_j$ (see Eq.~\ref{ee}) times $\mu_j$ (see Eq.~\ref{ansatz2}), hence proving that the vectors defined in Eq.~\ref{ansatz2} are, in fact, eigenvectors. 

Finally, according to the Perron-Frobenius theorem, for the zero eigenvalue with algebraic multiplicity equal to one, the corresponding eigenvector is $\mu_K=\mathbf{e}_N/\sqrt{N}$.
In this way we have obtained a complete basis of $N$ linearly independent eigenvectors related to the $N$ eigenvalues, namely they form an orthonormal basis, that is
\begin{equation}
	\mu_i\mu_j=\begin{cases} 1, & \mbox{if } i=j, \\ 0, & \mbox{if } i\neq j.
	\end{cases} \qquad\text{ and }\parallel\mu_i\parallel=1,\qquad\forall i=1,...,N.
\end{equation}
In particular, retaining the simple case $K=3$, the orthonormal basis of eigenvectors can be written as
\be
\mathbf{\mu}=\left( \begin{array}{cccccccc}
	\frac{1}{2\sqrt{2}} & \frac{1}{2\sqrt{2}} & \frac{1}{2} & 0 & \frac{1}{\sqrt{2}} & 0 & 0 & 0\\
	\frac{1}{2\sqrt{2}} & \frac{1}{2\sqrt{2}}  & \frac{1}{2}& 0 & -\frac{1}{\sqrt{2}}& 0 & 0 & 0\\
	\frac{1}{2\sqrt{2}} & \frac{1}{2\sqrt{2}}  & -\frac{1}{2}& 0 & 0 & \frac{1}{\sqrt{2}} & 0 & 0\\
	\frac{1}{2\sqrt{2}}& \frac{1}{2\sqrt{2}}  & -\frac{1}{2} & 0 & 0 & -\frac{1}{\sqrt{2}}& 0 & 0\\
	\frac{1}{2\sqrt{2}}& -\frac{1}{2\sqrt{2}} & 0 & \frac{1}{2} & 0 & 0 & \frac{1}{\sqrt{2}} & 0\\
	\frac{1}{2\sqrt{2}} & -\frac{1}{2\sqrt{2}}  & 0 & \frac{1}{2}& 0 & 0 & -\frac{1}{\sqrt{2}} & 0\\	
	\frac{1}{2\sqrt{2}} & -\frac{1}{2\sqrt{2}}  & 0 & -\frac{1}{2} & 0 & 0 & 0 & \frac{1}{\sqrt{2}}\\
	\frac{1}{2\sqrt{2}} & -\frac{1}{2\sqrt{2}}  & 0 & -\frac{1}{2}& 0 & 0 & 0 & -\frac{1}{\sqrt{2}}
 \end{array} \right).
\ee
The generalization to larger sizes is straightforward.

We conclude this section observing that
\be 
\sum_{i=1}^{N}\mu_{ik}=0,\forall k=2,...,N
\ee
and that
\begin{equation}
\mu_{ik}^2=\begin{cases} 1/2^{l+1},& \text{ if } i=2^{l+1}, \text{ with } l=0,...,K-1,\\
0 & \text{ if } i\neq 2^{l+1}, \text{ with } l=0,...,K-1.\label{propeig}
\end{cases} 
\end{equation}

\section*{Examples of applications \label{sec:applications}}
In this section we exploit the results found in the Sections ``Eigenvalues of the Dyson hierarchical graph'' and ``Eigenvectors'' to derive information about several processes which can be defined on the graph $\mathcal{G}$. These are just a few examples for illustrative purposes since, as underlined in the section ``The Laplacian spectrum and its applications'', the exact knowledge of the whole Laplacian spectrum can be applied in a very wide range of fields and situations.

\subsection*{Random Walks}
\begin{figure}[tb]
	\includegraphics[width=12cm]{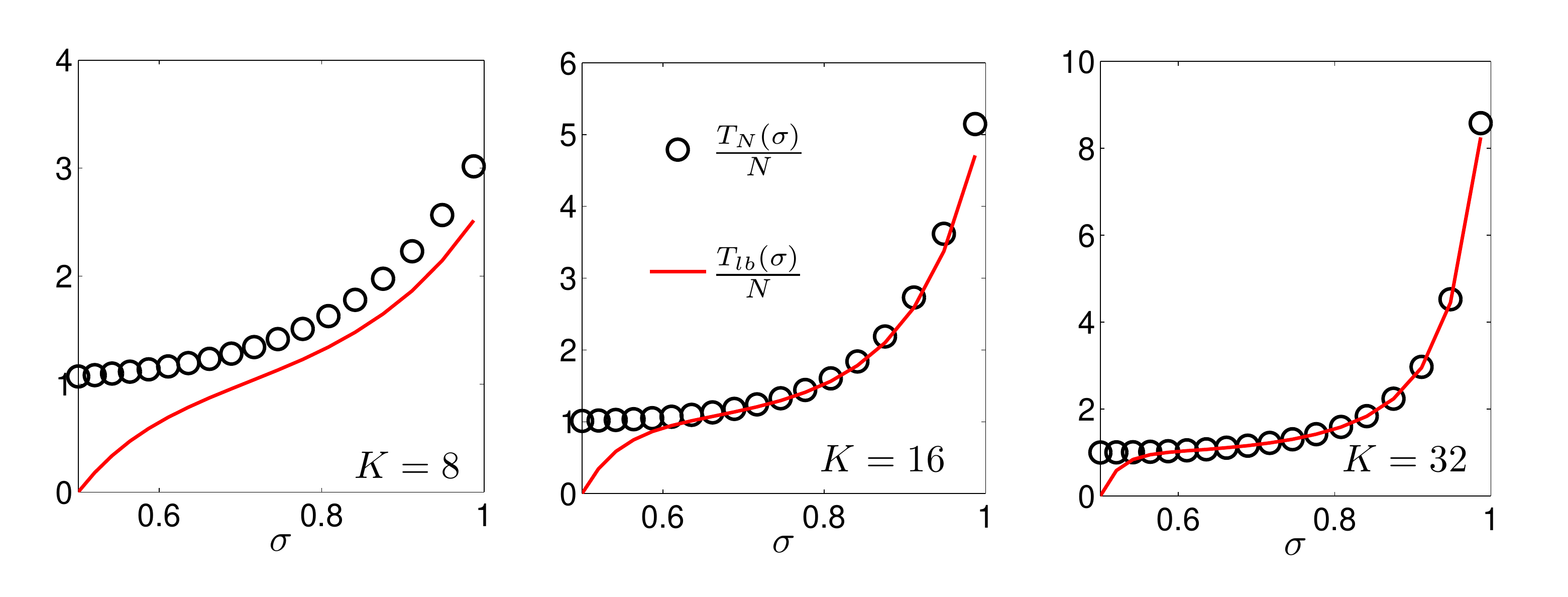}
	\caption{Normalized average mean-time of $T_j/N$ (circles) and its lower bound $T_{\text{lb}}/N$ (solid line) for $N=2^K$ with $K=8, 16, 32$. By plotting the normalized average mean time we get quantities that are roughly comparable as the size is varied. Left panel: for the relatively small size $N=2^8$ the lower bound $T_{\text{lb}}$ does not provide a quantitatively good estimate for $T_j$. Central panel: when the size is relatively large as $N=2^{16}$, the estimate is far improved. Right panel: when the size is very large as $N=2^{32} \approx 10^9$, $T_j \approx T_{\text{lb}}$ almost everywhere. }\label{bound}
\end{figure}

\begin{figure}[tb]
	\includegraphics[width=12cm]{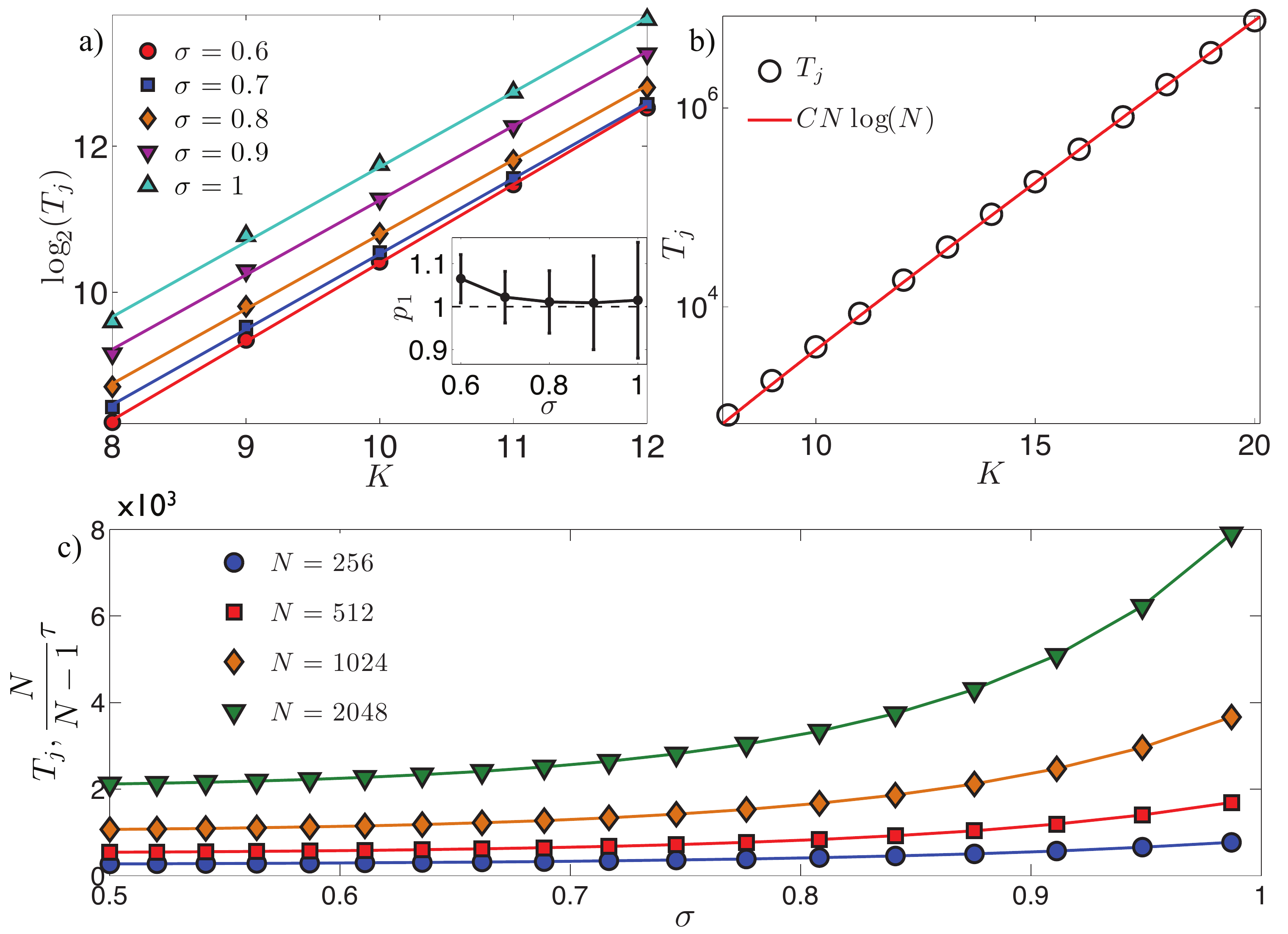}
	\caption{Panel $a$: Behavior of $\log_2(T_j)$ with respect to $K = \log_2(N)$. Different sets of data, corresponding to different choices of $\sigma$, are shown with different symbols, as explained by the legend. Data points are obtained by numerical evaluating Eq.~\ref{res}, while solid lines represent the best linear fit. In fact, as evidenced in Fig.~\ref{bound}, the lower bound given by $T_{\textrm{lb}}$ in Eq.~\ref{tlb} provides (in the limit of large size and for $\sigma$ far from its boundary values) a good approximation for $T_j$ and, in turn, in the limit of large size, $T_{\textrm{lb}}$ scales linearly with $N$ as shown in Eq.~\ref{eq:lb}. More precisely, in our fitting function $y = p_1 K + p_2$ we let the fit coefficients free and in the inset we compare the best-fit coefficient $p_1$ with the unitary value expected from Eq.~\ref{eq:lb}. The comparison is, within the error, satisfactory and further highlights that this approximation works better in the central region of the range $\sigma \in (1/2,1)$.
		Panel $b$: When $\sigma =1$ an exact expression for $T_j$ is achievable and here we compare the asymptotical analytical result, reported in Eq.~\ref{as}, with the numerical evaluation of $T_j$ from Eq.~\ref{res}, as the size $N=2^K$ varies, being $K\in[8,20]$; notice the semilogarithmic scale plot. One can see that the numerical evaluation of $T_j$ (circles) is well approximated by the function $CN\log_2(N)$, where $C =1/6$ (solid line).
		Panel $c$: The MFPT is plotted versus $\sigma$ and for different choices of the graph size, as indicated by the legend. 
		Data points are evaluated via the pseudo-inverse Laplacian method \cite{Fouss-2007IEEE} (in order to get a further check), while solid lines represent the numerical evaluation of Eq.~\ref{res}. Notice that, in the limit of large size, $T_j$ corresponds to the global  first passage time $\tau$.}\label{ht}
\end{figure}

A simple random walk embedded on an arbitrary graph $\mathcal{H}$ is characterized by the transition matrix $\mathbf{P} = \mathbf{W}^{-1} \mathbf{J}$, that is, the probability that, in a given time step, the walker jumps from $i$ to $j$ is $P_{ij} = J_{ij}/w_i$.
Now, from the complete knowledge of the Laplacian spectrum one can derive the dynamical properties of the random walker and, in particular, one can calculate the mean time taken by a random walker to \emph{first} reach a given node. This quantity plays a role in real situations such as transport in disordered media, neuron firing, spread of diseases and target search processes (see e.g., \cite{FirstPassage-Book2014,Redner-2001Book,Condamin-Nature2007}).
Without loss of generality, we can fix the target site on the node $j$ and focus on the \emph{mean first passage time} (MFPT) from node $i$  to node $j$, denoted as $T_{ij}$. 
According to the definition of MFPT for random
walks, we have $T_{jj} =  $0 and, for any $i \neq j$,
\be
T_{ij} = \sum_{k=1}^{N-1} P_{ik} T_{kj} +1,
\ee
 which can be rewritten in matrix form as
\be
\bar{\mathbf{T}} =  \bar{\mathbf{P}} \bar{\mathbf{T}} + \bar{\mathbf{e}} = \bar{\mathbf{W}}^{-1} \bar {\mathbf{J}} \bar{\mathbf{T}} + \bar{\mathbf{e}},
\ee
 where $\bar{\mathbf{e}}$  is the $(N - 1)$-dimensional unit vector $(1, 1, ..., 1)^T$; $\bar{\mathbf{T}}$ is the subvector of $\mathbf{T}$ obtained by deleting the $j$-th entry and whose $i$-th entry is $T_{ij}$ [Note: More precisely, the $i$-th entry of $\bar{\mathbf{T}}$ is $T_{ij}$ as long as $i<j$, while if $i>j$ the entry corresponding to $T_{ij}$ is the ($i-1$)-th one.]; 
$\bar{\mathbf{P}}$, $\bar{\mathbf{W}}$, and $\bar{\mathbf{J}}$  are, respectively, the submatrices of $\mathbf{P}$, $\mathbf{W}$, and $\mathbf{J}$ obtained
by deleting the $j$-th row and $j$-th column.
With some passages it is possible to express
$T_{ij}$ in terms of the spectra of $\mathbf{L}$  as (see \cite{Zhang-PRE2013b} for a detailed derivation)
\be \label{eq:Tij}
T_{ij} = \sum_{z=1}^{N} w_z \sum_{k=2}^N \frac{1}{\varphi_k} (\mu_{ik} \mu_{zk} - \mu_{ik} \mu_{jk} - \mu_{jk} \mu_{zk} + \mu_{jk}^2). 
\ee
%
%
%
%
Next, we can calculate the \emph{average mean time} $T_j$ to first reach the target node $j$, by averaging $T_{ij}$ over all possible starting sites, namely
\be
T_j \equiv \frac{1}{N-1} \sum_{i=1}^{N-1} T_{ij}.
\ee
By exploiting the result of Eq.~\ref{eq:Tij} we can recast the previous expression in terms of the Laplacian eigenvalues and eigenvectors as well, namely (see again \cite{Zhang-PRE2013b} for a detailed derivation)
\be \label{eq:Tj}
T_j = \frac{N}{N-1} \sum_{k=2}^N \frac{1}{\varphi_k}  \left (s \mu_{jk}^2 - \mu_{jk} \sum_{z=1}^N w_z \mu_{zk} \right),
\ee
where $s$ is the sum of the weighted degrees over all nodes, namely $s = \sum_{i=1}^N w_i = \sum_{i=1}^N \sum_{j=1}^N J_{ij}$. 
\newline
{In particular, for the graph $\mathcal{G}_K$ under study, we have that $s=Nw$, due to the homogeneity among nodes, and this also implies that $T_j$ is actually independent of $j$. Moreover, exploiting Eq.~\ref{propeig}, we can further simplify the general expression (\ref{eq:Tj}) as}
\begin{eqnarray}
T_j &=& \frac{N^2}{N-1}w\sum_{k=2}^N \frac{1}{\varphi_k}   \mu_{jk}^2\nonumber\\
&=& \frac{N^2}{2(N-1)}w(2^{2\sigma-1}-1)\sum_{l=0}^{K-1}\frac{2^{-l}}{2^{l(1-2\sigma)}-N^{1-2\sigma}}\label{sumval}\\
&=&\frac{N^2}{4(N-1)}\frac{2N^{1-2\sigma}(1-2^{2\sigma})+N^{-2\sigma}(2^{2\sigma}-2)+2^{2\sigma}}{2^{2\sigma}-1}\sum_{l=0}^{K-1}\frac{2^{-l}}{2^{l(1-2\sigma)}-N^{1-2\sigma}}.\label{res}
\end{eqnarray}
The sum appearing in Eq.~\ref{res} is not feasible for an explicit, general solution, yet, noticing that 
\be
\sum_{l=0}^{K-1}\frac{2^{-l}}{2^{l(1-2\sigma)}-N^{1-2\sigma}}\geq \sum\limits_{l=0}^{K-1}2^{l(2\sigma-2)}=\frac{1-N^{2\sigma-2}}{1-2^{2\sigma-2}},
\ee
we can provide $T_j$ with a lower bound, that is,
\be
T_j\geq \frac{N^2}{(N-1)}\frac{2N^{1-2\sigma}(1-2^{2\sigma})+N^{-2\sigma}(2^{2\sigma}-2)+2^{2\sigma}}{2^{2\sigma}-1}\frac{1 - N^{2\sigma-2}}{4 - 4^{\sigma}}=T_{\text{lb}}\qquad \text{ with }\sigma\in\Big(\frac{1}{2},1\Big).\label{tlb}
\ee
The reliability of this bound is shown in Fig.~\ref{bound}: by varying $K$ and $\sigma$ we see that the bound is more accurate when the size is large (i.e., $K \gg 1$) and when the pattern of weights is less homogeneous (i.e., $\sigma$ far from $1/2$). 
The leading term of $T_{\textrm{lb}}$ scales as 
\be \label{eq:lb}
T_{\textrm{lb}} \approx \frac{2^{2\sigma}}{(2^{2\sigma}-1)(4-2^{2\sigma})} \times N \left(1 - \frac{1}{N^{2 \sigma -2}} \right), 
\ee
which evidences that (at least for large size and for $\sigma$ far from its boundary values) the MFPT is well approximated by a linear growth with the size $N$, as corroborated by Fig.~\ref{ht} (panel $a$).
Moreover, one can show that (see also Fig.~\ref{bound})
\be
T_j \leq T_j(\sigma=1)
\ee
and, for $\sigma=1$, we are able to compute explicitly the sum in Eq.~\ref{sumval}. In fact, fixing $\sigma=1$, it becomes
\be
\sum_{l=0}^{K-1}\frac{2^{-l}}{2^{-l}-N^{-1}}=-\frac{1}{2}+\frac{2(3N-1)}{(4N-1)(1-2N)}+K.
\ee
Plugging this expression in Eq.~\ref{res}, and posing $\sigma=1$, we obtain the related exact value of $T_j$ as
\be
T_j=\frac{1}{6}\Big(N-\frac{1}{2}\Big)\Big[K+\frac{2(3N-1)}{(4N-1)(1-2N)}-\frac{1}{2}
\Big]\approx CN\log_2(N),\qquad\text{ as }N\rightarrow +\infty,\text{ when }\sigma=1,\label{as}
\ee
where in the last approximation we outlined the leading term being {$C = 1/6$} (see also Fig.~\ref{ht}, panel $b$).

Therefore, by comparing the approximate result found for $\sigma \in (1/2,1)$ (see Eq.~\ref{eq:lb}) and the exact result found for $\sigma=1$ (see Eq.~\ref{as}), we expect that $T_j$ depends functionally on $\sigma$.

In the case the location of the target is unknown, one can still obtain a characteristic time by averaging $T_j$ over all possible target locations, hence getting the so-called \emph{global first passage time}, referred to as $\tau$ and defined as 
\be \label{eq:34Z}
\tau  \equiv \frac{1}{N(N-1)} \sum_{j=1}^N \sum_{\substack{i=1\\i \neq j}}^N T_{ij} = \frac{1}{N} \sum_{j=1}^N T_j.
\ee
By plugging Eq.~\ref{eq:Tj} into Eq.~\ref{eq:34Z}, we get
\be 
\tau = w \sum_{k=2}^N \frac{1}{\varphi_k},\label{t}
\ee
which can be further developed by exploiting the previous results on eigenvalues (see Eq.~\ref{ee} and \ref{e}).
However, for the current graph, exploiting the homogeneity among the nodes, one can immediately write
\be
\tau = \frac{1}{N} \sum_{j=1}^N T_j = \frac{N-1}{N} T_j \approx T_j, \label{HT}
\ee
where the last passage holds for large $N$ (see Fig.~\ref{ht}, panel $c$).

To summarize, the MFPT $T_j(\sigma)$ on $\mathcal{G}$ can be bounded as
\be
\mathcal{O}(N) \approx T_{\textrm{lb}} \leq T_j(\sigma) \leq T_j(\sigma=1) \approx \mathcal{O}(N \log N).
\ee
In general, as expected, $T_j(\sigma)$ grows with $\sigma$ since, when the pattern of weights is more heterogeneous (i.e., $\sigma$ is large), the graph tends to get disconnected, the mixing time is large and reaching the farthest nodes takes more and more time.\\
{The leading asymptotic dependence on the network size displayed by $T_j(\sigma)$ can be compared to the scalings found for other structures displaying extensive connectivity (i.e., the coordination number of a subset of nodes diverges with the network size). 
For instance, for the class of pseudofractal networks studied in \cite{Zhang-PRE2009}, the MFPT scales sublinearly with the size $N$ as long as the target node is central, while it scales linearly when the target node is peripheral; analogous results hold for the deterministic scale-free networks studied in \cite{Agliari-PRE2009}. Interestingly, a scaling $\sim N \log N$ was evidenced for the global MFPT in the small-world exponential treelike networks studied in \cite{Zhang-JStat2011}, where a modular topology with loosely connections between different modules is also responsible for preventing the walker from exploring fluently the underlying space. }

\subsection*{Quantum Walks}
As stressed above, random walks constitute a basic example of dynamical process affected by the underlying topology. Their quantum-mechanical version, i.e. the quantum walks, constitute an advanced tool for building quantum algorithms and for modeling the coherent transport of mass, charge or energy (see e.g., \cite{Kempe-ContPhys2003, MB-PhysRep2011,Venegas-QIPJ2012}), and display as well a strong dependence on the topological properties of the embedding structure and, accordingly, much of their properties can be expressed in terms of the Laplacian spectrum. Before deepening this point it is worth summarizing briefly a few definitions (here we just focus on the continuous time version of quantum walks, i.e., continuous-time quantum walks), while we refer to \cite{MB-PhysRep2011} for an extensive review.
\newline
In such a quantum-mechanical context, the transport has to be formulated in Hilbert
space and one assumes that the states $| j \rangle$  representing the
nodes span the whole accessible Hilbert space; also, it
is assumed that the states are orthonormal and complete, i.e.,
$\langle i | j \rangle = \delta_{ij}, \textrm{and} \sum_{i} | i \rangle \langle i | = 1$.
The behavior of the walker can be described by the transition amplitude $\alpha_{k,j}(t)$ from state $| j \rangle$ to
state $| k \rangle$, which obeys the following Schr\"{o}dinger equation\footnote{The constant $\hbar$ is set equal to $1$.}
\begin{equation}\label{eq:schrodinger}
\frac{d}{dt} \alpha_{k,j}(t)=-i \sum_{l=1}^{N} H_{kl} \alpha_{l,j}(t),
\end{equation}
where $\mathbf{H}$ is the quantum-mechanical Hamiltonian associated to the process and it can be identified with the classical transfer matrix, i.e., $\mathbf{H} \equiv  \mathbf{L}$ \cite{Farhi-PRA1998,MB-PhysRep2011}.
The squared magnitude of the transition amplitude provides the quantum mechanical
transition probability $\pi_{k,j}(t) \equiv |\alpha_{k,j}(t)|^2$.

The performance of the continuous-time quantum walk (CTQW) can be estimated in terms of the return probabilities $\pi_{j,j} (t)$: a quick temporal decay of these probabilities implies a fast transport
through the network. In order to make a global statement
on the performance, one considers the average return probability $\bar{\pi}$ given by \cite{MB-PhysRep2011}
\be
\bar{\pi}(t) = \frac{1} {N} \sum_{j} \pi_{j,j}(t) \geq   \left| \frac{1}{N} \sum_{j=1}^N \alpha_{j,j}(t) \right|^2 = \left| \frac{1}{N} \sum_{n=}^N e^{-i \lambda_n t}  \langle \mu_n | \sum_j | j \rangle \langle j| \mu_n \rangle \right |^2 = \left| \frac{1}{N} \sum_{n=1}^N e^{-i \lambda_n t} \right|^2,
\ee
where in the last passage we exploited the Cauchy-Schwarz inequality to get a lower bound for
$\bar{\pi}(t)$ which does not depend on the eigenvectors.

For classical diffusion, the long-time limit of the
transition probabilities reaches the equipartition value $1/N$. In
contrast, due to the unitary time evolution, for CTQW neither $\pi_{i,j}(t)$ nor $\bar{\pi}(t)$  decay to a given
value at long times, but rather oscillate around the corresponding
long-time average $\bar{\chi}$ which, for $\bar{\pi}(t)$, is given by \cite{MB-PhysRep2011}
\be
\bar{\chi} \equiv \lim_{T \rightarrow \infty} \frac{1}{T} \int_0^{T} \bar{\pi}(t) dt = \frac{1}{N} \sum_{n,m} \delta_{\lambda_n, \lambda_m} |\langle  j | \mu_n \rangle |^2 | \langle  j | \mu_m \rangle |^2.
\ee
Remarkably, one can obtain a lower bound which does not depend on
the eigenvectors and reads as \cite{MB-PhysRep2011,KDM-PRL2015}
\be \label{lta}
\bar{\chi}  \geq \frac{1}{N^2} \sum_{n,m} \delta_{\lambda_n, \lambda_m} \equiv \bar{\chi}_{\textrm{lb}};
\ee
{namely, $\bar{\chi}_{\textrm{lb}}$ is the sum of the squares of the normalized multiplicities.}
Therefore, the full knowledge of the Laplacian eigenvalue spectrum allows estimating (at least via bounds) whether the quantum transport on a given structure propagates relatively fast.
Notice that, actually, $\bar{\chi}_{\textrm{lb}}$ just depends on the multiplicities of the distinct eigenvalues.

Indeed, for quantum transport processes the degeneracy of
the eigenvalues plays an important role, as the differences
between eigenvalues determine the temporal behaviour, while for
classical transport the long time behaviour is dominated by the
smallest eigenvalue. Situations where only a few, highly-degenerate eigenvalues are present are related to slow CTQW
dynamics, while when all eigenvalues are non-degenerate the
transport turns out to be efficient (i.e., it spreads out fast) \cite{MB-PhysRep2011}.

However, in real-systems an excitation typically undergoes decay (e.g., by exciton recombination) or absorption (e.g., at the reaction center of light-harvesting antenna) and, in such cases, the total probability to find the excitation within
the network is not conserved. In order to keep track of such loss processes we can insert an absorption term in the Hamiltonian $\mathbf{H}$ appearing in Eq.~\ref{eq:schrodinger}. Let us consider the case where the excitation can only vanish at certain
nodes (called ``traps''), making up a set $\mathcal{M}$ with cardinality $|\mathcal{M}| = M$. 
Then, the new Hamiltonian reads as $\mathbf{H}^{\prime} = \mathbf{H} - i \mathbf{\Gamma}$, where $\mathbf{\Gamma}$ can be written as a diagonal matrix with elements $\Gamma_{jj} = \Gamma$ as long as $j \in \mathcal{M}$ and zero otherwise; $\Gamma$ represents the (tunable) absorbing rate.

As long as the absorbing rate $\Gamma$ is small, i.e., $\Gamma \ll 1$, this problem can be treated perturbatively getting that the mean survival probability $\Pi_M(t)$ can be approximated by a sum of exponentially decaying terms:
\begin{equation}\label{eq:pi_asym}
\Pi_{M}(t) \approx \frac{1}{N-M} \sum_{l=1}^{N} e^{-2 \gamma_l t},
\end{equation}
where $M$ is the number of traps and $\gamma_l$ is the imaginary part of the eigenvalue $\varphi_l^{\prime}$ of the perturbed Hamiltonian $\mathbf{H}^{\prime}$ (with approximation at first order), namely $\gamma_l = \Gamma \sum_{m \in \mathcal{M}} \left| \langle m | \mu_l \rangle \right|^2$ \cite{MB-PhysRep2011,A-PHYSA2011}.

Let us now resume the hierarchical graph $\mathcal{G}_K$ and exploit the results obtained in Sections ``Eigenvalues of the Dyson hierarchical graph'' and ``Eigenvectors'' to derive an estimate for $\bar{\chi}$ and for $\Pi_{M}(t)$.

We start with $\bar{\chi}$ and notice that the estimate in Eq.~\ref{lta} only accounts for the degeneracy of the eigenvalues which, as can be seen in Eqs.~\ref{ee}-\ref{e}, is not affected by $\sigma$, but only depends on the system size. As a result, $\chi_{\text{lb}}$ does not depend on $\sigma$. Moreover, exploiting the expression for $\varphi_k$ obtained in Eqs.~\ref{ee}-\ref{e}, we can get that $\chi_{\text{lb}}$ for a generic network of size $N=2^K$ reads as 
 \begin{equation}
 	\chi_{\text{lb}}=\frac{N^2+2}{3N^2} \xrightarrow[K \rightarrow \infty]{} \frac{1}{3}.\label{rec}
 \end{equation}

We can therefore derive that, no matter the system size and the value of $\sigma$, the long time average $\bar{\chi}$ is always finite and larger than $1/3$. This means that the coherent transport on $\mathcal{G}_K$ tends to get stuck nearby the starting node. Of course, this stems from the fact that couplings decay exponentially fast with the distance among nodes.
Actually, localization phenomena also occurs for the classical diffusion, where, as the size $N$ gets larger and larger the mixing time grows exponentially and ergodicity breaks down in the limit $N \rightarrow \infty$ \cite{TA-PRE2016,ABGGTT-PRL2015,ABGGTT-PRE2015}. However, at finite size, for classical propagation the equipartition is eventually reached, while for the coherent propagation the localization effect is significant already at small size.

Let us now move to $\Pi_{M}(t)$; according to Eq.~\ref{eq:pi_asym} it is possible to get a long time estimate in terms of the imaginary part $\gamma_l$ of the eigenvalues of the perturbed Hamiltonian $\mathbf{H'}$. This can be calculated as follows:
\be
\gamma_l = \langle \mu_l | \mathbf{\Gamma} | \mu_l \rangle = \Gamma \sum_{i \in \mathcal{M}}  [\mu_{il}]^2 = \Gamma \frac{2^{g_l}}{N} \sum_{i \in \mathcal{M}}  (1 - \delta_{\mu_{il},0})
\ee
where in the second passage we exploited the diagonal form of the matrix $\mathbf{\Gamma}$ and in the third passage we exploited the results found in Section ``Eigenvectors''. More precisely, $\mathbf{g}$ is a vector, whose entry $g_l$ corresponds to the number of null entries in the eigenvector $\mu_l$, namely $\mathbf{g} = (0, 0, 1, 1, 2, 2, 2, 2, 3, ...., N-2)$, and $\delta_{x,y}$ is the Kronecker delta in such a way that the summation just returns the number of non-null entries in the eigenvector $\psi^{(l)}$ that match a trap position.
For instance, when $M=1$, one finds 
\begin{eqnarray}
\nonumber
\gamma_1 &=& 1/N,\\ 
\nonumber
\gamma_2 &=& 1/N,\\ 
\nonumber
\gamma_3 &=& 2/N, ~ \gamma_4 = 0,\\
\nonumber
\gamma_5 &=& 4/N, ~ \gamma_6 = \gamma_7 =\gamma_8 =0,\\
\nonumber
\gamma_9 &=& 8/N , ~\hdots ,\\
\nonumber
\vdots
\end{eqnarray}
or any other outcome obtained under the permutation of the elements pertaining to the same line (e.g., $\gamma_6 =4/N,  \gamma_5 = \gamma_7 =\gamma_8 =0$).
In general, in the presence of an arbitrary number $M$ ($M \leq N$) of traps, we have $\gamma_1 = \gamma=2 = M/N$, while for $l>2$ the value of $\gamma_l$ depends on the trap arrangement. Moreover, the smallest eigenvalue is null and the second smallest eigenvalue is $M/N$, whose multiplicity grows with $M$. 
Each null eigenvalue corresponds to a block of $\mathcal{G}_K$ which is trap free and this yields to a contribute to $\Pi_M(t)$ which survives even at long times. 
Therefore, the trap arrangement which maximizes the survival probability is the one where traps are set as close as possible each other, while the trap arrangement which minimizes the survival probability is the one where traps are scattered broadly. This is rather intuitive as the quantum walk tends to remain localized around its starting point in such a way that if traps are assembled within a block, the walker can avoid them as there exist eigenstates living on different blocks, while if each block (even the smallest one, i.e. the dimer) is occupied by at least one trap the quantum walk can not avoid them \cite{Agliari-IJBC2010}.

\subsection*{Dynamics of polymer networks under external forces}

Another field where the Laplacian spectrum is extensively exploited is polymer physics and, in particular, as for the investigation of the relationship between the topology of a polymer and its dynamics. To this purpose, the so-called generalized Gaussian structures are conveniently exploited. 
These are a generalization of the Rouse model \cite{Rouse-JCP1953,Gurtovenko-AdvPolymSci2005}, meant for linear polymer chains, to systems with arbitrary topology. The polymer is modeled as a structure consisting of $N$ beads (each bead corresponding to a node) connected by harmonic springs (we refer to \cite{Gurtovenko-AdvPolymSci2005} for a discussion of the underlying assumptions). 
\newline
Understanding how the underlying geometries of polymeric materials affect their dynamic behavior is becoming of increased importance as new polymeric materials with more and more complex architectures are synthesized \cite{Gurtovenko-AdvPolymSci2005}.

Let us consider a set of $N$ beads, all  subject to the same friction constant $\zeta$ with respect to the sorrounding viscous medium, and connected pairwise by harmonic strings. The pattern of the related string constants is provided by the matrix $\mathbf{L}$ in such a way that for the couple $(i,j)$ it reads  $C L_{ij}$, where $C$ is a suitable constant. The equation describing the motion of the $j$-th bead is the following Langevin equation
\begin{equation}
	\zeta\frac{d \textbf{R}_j(t)}{dt}+C\sum_{i=1}^{N}L_{ji} \textbf{R}_i(t)= \textbf{f}_j(t)+ \textbf{F}_j(t),
\end{equation}
where $\textbf{R}_j(t)=(X_j(t),Y_j(t),Z_j(t))$ represents the position in the space of the $j$-th bead at time $t$, $\mathbf{L}$ is the Laplacian matrix associated to the structure of the polymer, $\mathbf{f}_j(t)$ is the thermal Gaussian noise [Note: More precisely, it is assumed that $\langle f_j (t)\rangle =0$ and $\langle f_{j \alpha}(t)f_{i \beta}(t^{\prime})\rangle=2 k_B T \zeta \delta_{\alpha \beta}\delta_{ij}\delta(t-t^{\prime})$, where $k_B$ is the Boltzmann constant, $T$ is the temperature, {$\zeta$ is the friction constant}, $\alpha$ and $\beta$\ represent the $x$, $y$, and $z$ directions.], and $\textbf{F}_j(t)$ is the sum of all external forces acting on the $j$-th bead.  
\newline
One can show that, acting (from time $t = 0$ onwards) with a constant external
force $F_k(t) = F \delta_{jk}$ on a single bead $j$ of the polymer
in the, say, $y$ direction, the mean displacement of a bead turns out to be \cite{Gurtovenko-AdvPolymSci2005}
\be\label{eq:constF}
\langle Y(t) \rangle = \frac{Ft}{N \zeta} +  \frac{F}{\gamma N \zeta} \sum_{i=2}^{N} \frac{1 - e^{-\gamma \varphi_i t}}{\varphi_i}
\ee
where $\gamma = C/\zeta$ is the bond rate constant and $\{ \varphi_i \}_{i=2}^N$ are the non-null eigenvalues of
matrix $\mathbf{L}$, with $\varphi_1$ being the unique (for otherwise the polymer would be split in several parts) smallest eigenvalue $0$. 

When the applied force is harmonic, namely $F_k(t) = F e^{i \omega t} Y_{k}(t)$, say along the $x$ direction, the related response function is the so-called complex dynamic (shear) modulus $G^*(\omega)$, whose real, i.e. $G^{\prime}(\omega)$, and imaginary, i.e., $G^{\prime \prime}(\omega)$, components are also referred to as the storage and the loss moduli, respectively \cite{Gurtovenko-AdvPolymSci2005}.  In the generalized Gaussian structure model these are given by 
\begin{eqnarray}
\label{eq:loss1}
G^{\prime}(\omega)=\nu k_BT\frac{1}{N}\sum_{i=2}^{N}\frac{\omega^2}{\omega^2+(2\gamma\varphi_i)^2},\\
\label{eq:loss2}
G^{\prime\prime}(\omega)=\nu k_BT\frac{1}{N}\sum_{i=2}^{N}\frac{2\gamma\varphi_i \omega}{\omega^2+(2\gamma\varphi_i)^2},
\end{eqnarray}
where $\nu$ represents the beads per unit volume.

\begin{figure}[h!]
	\includegraphics[width=15cm]{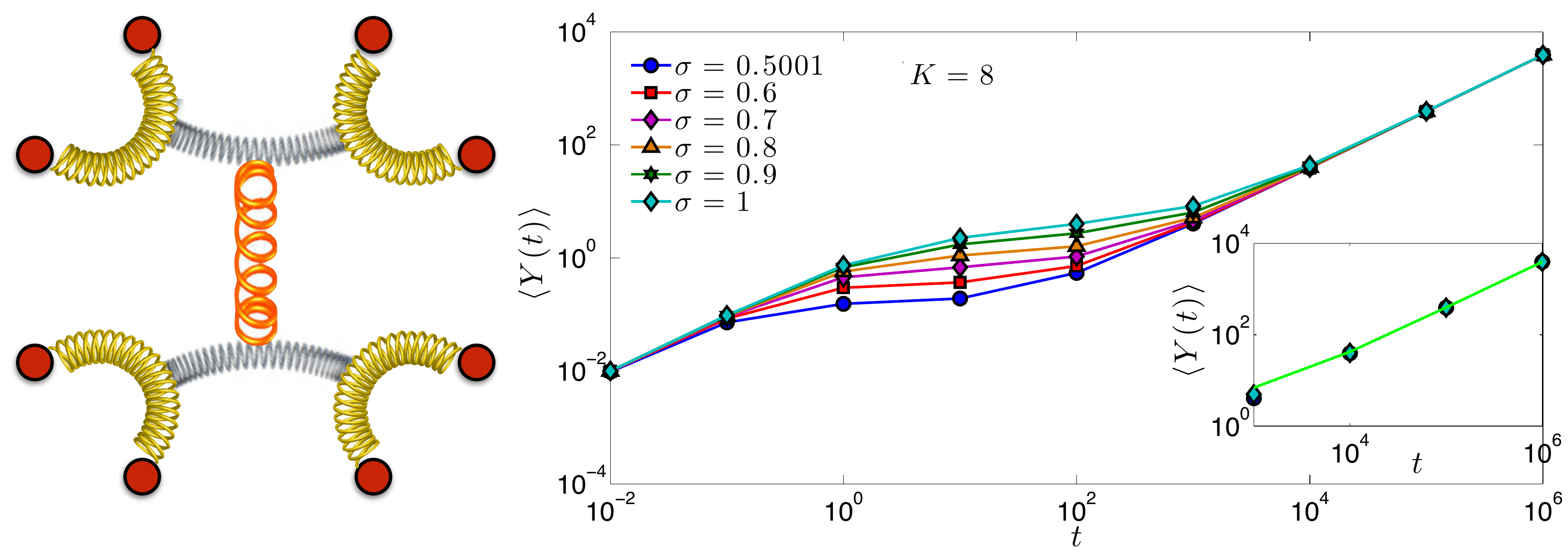} \label{fig:spring}
	\caption{Left panel: Schematic representation of the polymer network described by $\mathcal{G}_K$: $N=8$ beads are pairwise coupled via couplings whose pattern exhibits a hierarchical structure. Right panel: Trend of the mean displacement of a bead $\langle Y(t)\rangle$ as a function of $t \in[10^{-2},10^6]$, for $K=8$ and for different values of $\sigma\in(1/2,1]$, as shown by the legend. As highlighted analytically (see Eqs.~\ref{zero} and \ref{inf}), for short and long times the mean displacement grows linearly with time and the rate of growth is independent of $\sigma$. On the other hand, in the intermediate time interval, the rate of growth is sub-linear and slower when the pattern of couplings is more homogeneous. The data points shown here are obtained by numerically evaluating Eq.~\ref{constf2}, where we set $F/\zeta=1$ and $\gamma=2$, lines are guide to eye. The inset shows a zoom on the long-time region where the two extremal cases $\sigma=0.5001$ and $\sigma=1$ are compared with the linear law $y = t/N + \tau/(Nw)$ (see Eqs.~\ref{inf}, solid line).} \label{fig:Y}
\end{figure}

Let us now resume the graph $\mathcal{G}_K$: 
as discussed previously, this represents a set of interconnected nodes where the coupling pattern exhibits hierarchy and modularity (see Fig.~\ref{fig:Y}, right panel).

\noindent
Exploiting the results found for the hierarchical graph under study (see Eqs.~\ref{ee}-\ref{e}), we can restate Eq.~\ref{eq:constF} as
\be \label{constf2}
\langle Y(t) \rangle = \frac{Ft}{N \zeta} +  \frac{F}{2 \gamma \zeta} \sum_{n=0}^{K-1} \frac{1 - e^{-\gamma \varphi_n t} }{2^n \varphi_n}.
\ee
In the limit of very short time, the terms to be summed up in Eq.~\ref{constf2} can be approximated as $(1 - e^{-\gamma \varphi_n t})/(2^n\varphi_n) \approx \gamma  t  2^{-n} $, hence obtaining the linear scaling 
\be 
\langle Y(t) \rangle {\underset{t \to 0}\approx} \frac{Ft}{\zeta}.
\label{zero}
\ee 
Similarly, in the limit of very long times, we can write $(1 - e^{-\gamma \varphi_n t})/(2^n\varphi_n) \approx 2^{-n} / \varphi_n$ and, recalling the results in Eq.~\ref{t}, we get 
\begin{equation}
	\langle Y(t)\rangle {\underset{t \to \infty}\approx} \frac{Ft}{N\zeta} \left( 1 +\frac{\tau}{\gamma w} \right).\label{inf}
\end{equation}
{Notice that the last term in the right hand side is proportional to the radius of gyration of the polymer considered \cite{SB-JPA1995}.}\\
The results collected in Eqs.~\ref{zero} and \ref{inf} show that, in the limit of short and long times the mean displacement varies linearly with time, with a rate which is independent of $\sigma$. Therefore, the most interesting regime is the intermediate one, where the pattern of weights can possibly play a role. In fact, as shown in Fig.~\ref{fig:Y}, this is the case and the resulting displacement is enhanced when the network is more inhomogeneous (i.e., $\sigma$ is large).

\begin{figure}[h!]
	\includegraphics[width=14cm]{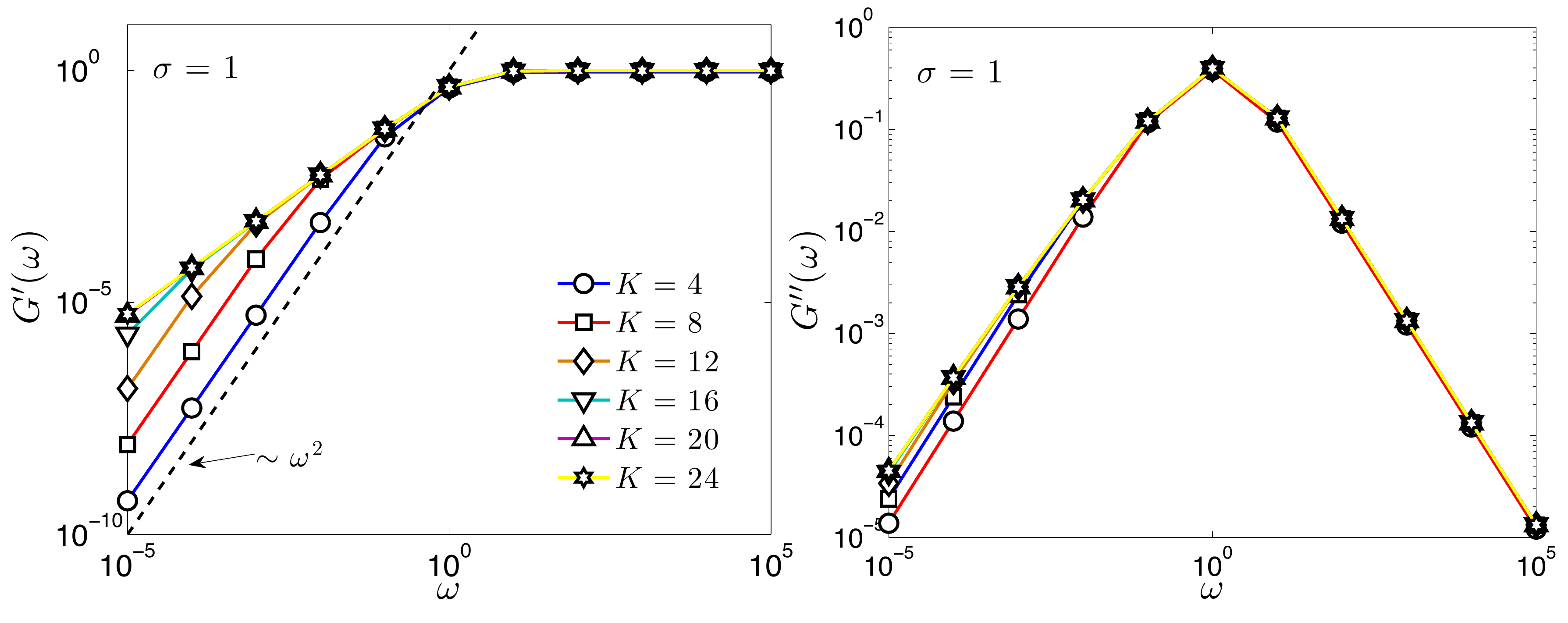}
	\caption{Gain $G'$ (left panel) and loss $G''$ (right panel) terms versus $\omega\in[10^{-5},10^5]$ in a logarithmic scale, for $\sigma=1$ and for a network of size $N=2^K$ with $K=4, 8, 12, 16, 20, 24$, as explained by the common legend. As expected, the gain term $G'$ is monotonically increasing with $\omega$, while the loss term exhibits a peak at intermediate frequencies. Here data are obtained by numerically evaluating Eqs.~\ref{gainterm} and \ref{lossterm}, where we set $\gamma =1$ and $\nu k_B T=1$; solid lines are guides to the eye.}\label{gainloss}
\end{figure}

As for the storage and loss moduli, exploiting the results of Section \textquotedblleft Eigenvalues of the Dyson hierarchical graph\textquotedblright, Eqs.~\ref{eq:loss1} and \ref{eq:loss2} can be recast as
\begin{eqnarray}
 G^{\prime}(\omega)  
 &=& \frac{\nu k_BT}{2} (2^{2\sigma-1}-1)^2\sum_{n=0}^{K-1}\frac{1}{2^n}\frac{\omega^2}{[\omega(2^{2\sigma-1}-1)]^2+4\gamma^2(2^{n(1-2\sigma)}-N^{1-2\sigma})^2},\label{gainterm}\\
 G^{\prime \prime}(\omega)
 &=&  \nu k_BT(2^{2\sigma-1}-1) \sum_{n=0}^{K-1}\frac{2^{-n}(2^{n(1-2\sigma)}-N^{1-2\sigma}) \omega}{(2^{2\sigma-1}-1)^2\omega^2+4(2^{n(1-2\sigma)}-N^{1-2\sigma})^2}\label{lossterm}.
\end{eqnarray}
The behavior of $G^{\prime}$ and of $G^{\prime\prime}$ has been evaluated numerically for $\sigma$ spanning in $(1/2,1]$; the results for the case $\sigma=1$ are shown in Fig.~\ref{gainloss}.
As expected, $G^{\prime}$ behaves as $\omega^2$ at low frequencies and as $\omega^0$ at high frequencies.
In the intermediate regime the growth of $G^{\prime}$ is smoothened and the width of such an intermediate scaling increases with $\sigma$. A similar behavior was evidenced for other unweighted networks displaying small-world features \cite{Zhang-SciRep2015,Liu-JCP2013}.
As for $G^{\prime \prime}$,  it peaks at around $\omega \approx 1$, the exact value depending on the choice of $\sigma$. 
For relatively small frequencies, $G^{\prime \prime}$ is larger when the pattern of weights is more inhomogeneous (i.e., $\sigma$ large), while for relatively large frequencies $G^{\prime \prime}$ is larger when the pattern of weights is more homogeneous (i.e., $\sigma$ small). In any case, the expected scalings $G^{\prime \prime}(\omega) \sim \omega$ and $G^{\prime \prime}(\omega) \sim \omega^{-1}$ are recovered. This is again consistent with the results found for small-world structures in \cite{Zhang-SciRep2015,Liu-JCP2013}.

\begin{figure}[h!]
	\includegraphics[width=10cm]{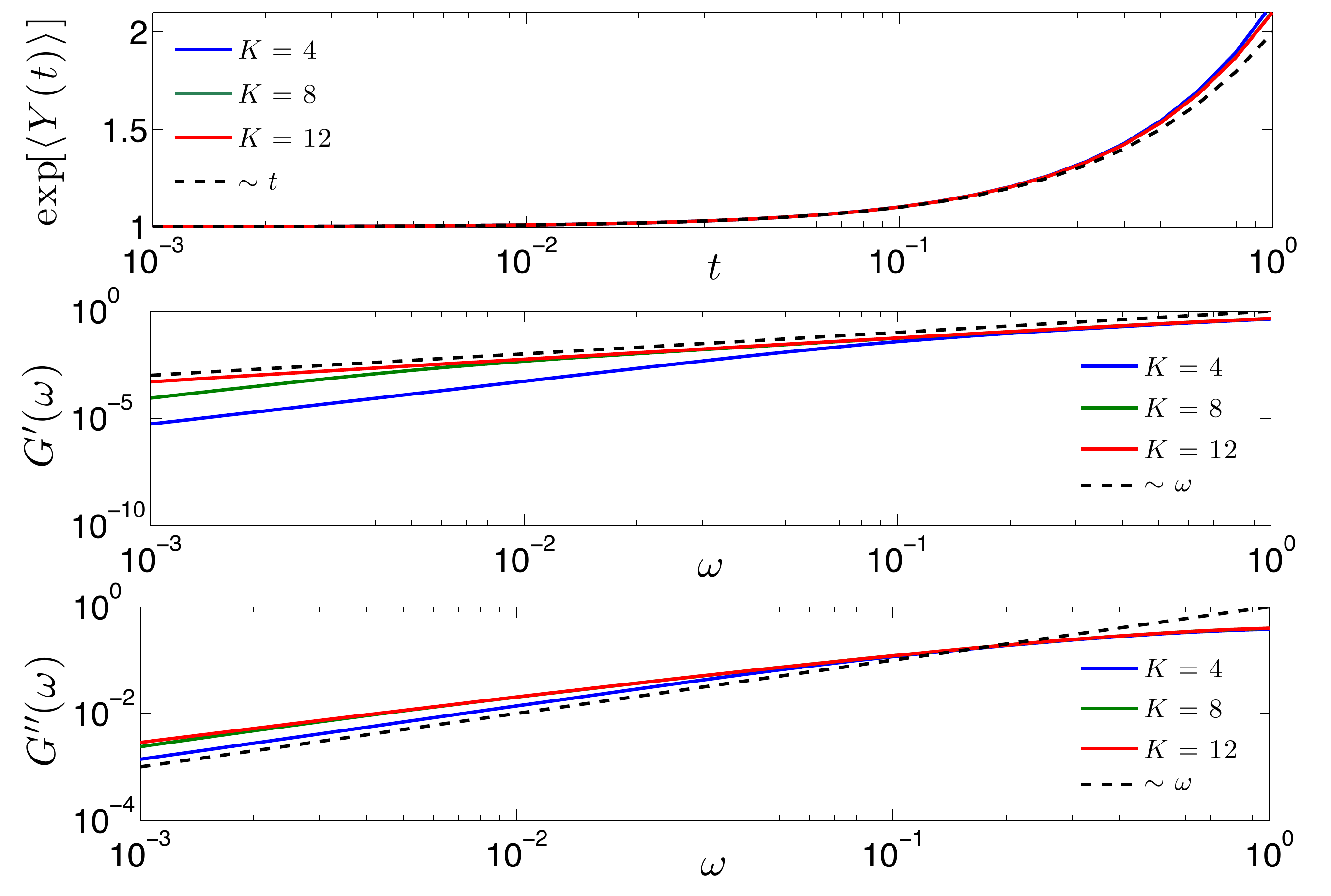}
	\caption{{These panels focus on the intermediate time/frequency domain for the mean displacement $\langle Y(t) \rangle$ (upper panel), for the storage modulus $G^{\prime}$ (middle panel) and for the loss modulus $G^{\prime \prime}$ (lower panel), calculated for $\sigma=1$; notice the logarithmic scale on the $x$-axis. Three different sizes ($N=2^K$ with $K=4, 8, 12$) are compared and the theoretical expected behavior is also shown (dashed line). In particular, for the mean displacement the expected logarithmic scaling is highlighted by plotting the (natural) logarithm of $\langle Y(t) \rangle$ and comparing it with a linear trend.}}\label{fig:intermediate}
\end{figure}

For the particular case $\sigma=1$ we can estimate the asymptotic behaviour of $G'$ by upper-bounding the expression in Eq.~\ref{gainterm}, that is
\begin{eqnarray}
G^{\prime}(\omega)|_{\sigma=1} &=& \frac{\nu k_BT \omega^2}{2}\sum_{n=0}^{K-1}\frac{2^{-n}}{\omega^2+4\gamma^2(2^{-n}-2^{-K})^2}\label{eq:prima}\\
&\leq & \frac{N\nu k_BT \omega^2}{8}\sum_{n=0}^{K-1}\frac{2^n}{(2^n-1)^2}\nonumber\\
&=& \frac{N\nu k_BT \omega^2}{8}\left[ \log 2 + \frac{2}{\log 2}+ \frac{N(6N^2-8N+3)}{2(N^2-3N+1)^2}\right].
\end{eqnarray}
{The previous equation shows that $G^{\prime}(\omega)|_{\sigma=1}$ scales at most linearly with $N$. Moreover, in the limit of low frequencies ($\omega \ll \gamma/N$), from Eq.~\ref{eq:prima} we can get $G^{\prime}(\omega)|_{\sigma=1} \sim \omega^2 N$; this scaling is corroborated in Fig.~\ref{gainloss} (left panel), where one can also see that this regime in the $\omega$ space shrinks with the size. In the opposite limit of high frequencies $(\omega \gg 2 \gamma)$, from Eq.~\ref{eq:prima} we get $G^{\prime}(\omega)|_{\sigma=1} \sim \omega^0 N^0$; again, the scaling is corroborated in Fig.~\ref{gainloss} (left panel) and, this time, the width of this regime in the $\omega$ space does not depend on $N$.\\
Finally, as discussed in the section ``Eigenvalues of the Dyson hierarchical graph'', the case $\sigma =1$ corresponds to $d_s=2$ which is a critical value over which the polymeric structures collapse (however, under external forces, such polymers can unfold \cite{Schiessel-PRE1998}, which makes this analysis reasonable). For the case $d_s=2$ the observables studied in this section display characteristic asymptotic behaviors in the intermediate time/frequency domain (see \cite{Zhang-SciRep2015} and references therein), namely $\langle Y(t) \rangle \sim \log(t)$ and $G^{\prime}(\omega) \sim G^{\prime \prime}(\omega) \sim \omega$. These scalings have been successfully checked also for the graph $\mathcal{G}$ under study, as shown in Fig.~\ref{fig:intermediate}.}

 \subsection*{Modularity}
\begin{figure}[tb]
	\includegraphics[width=14cm]{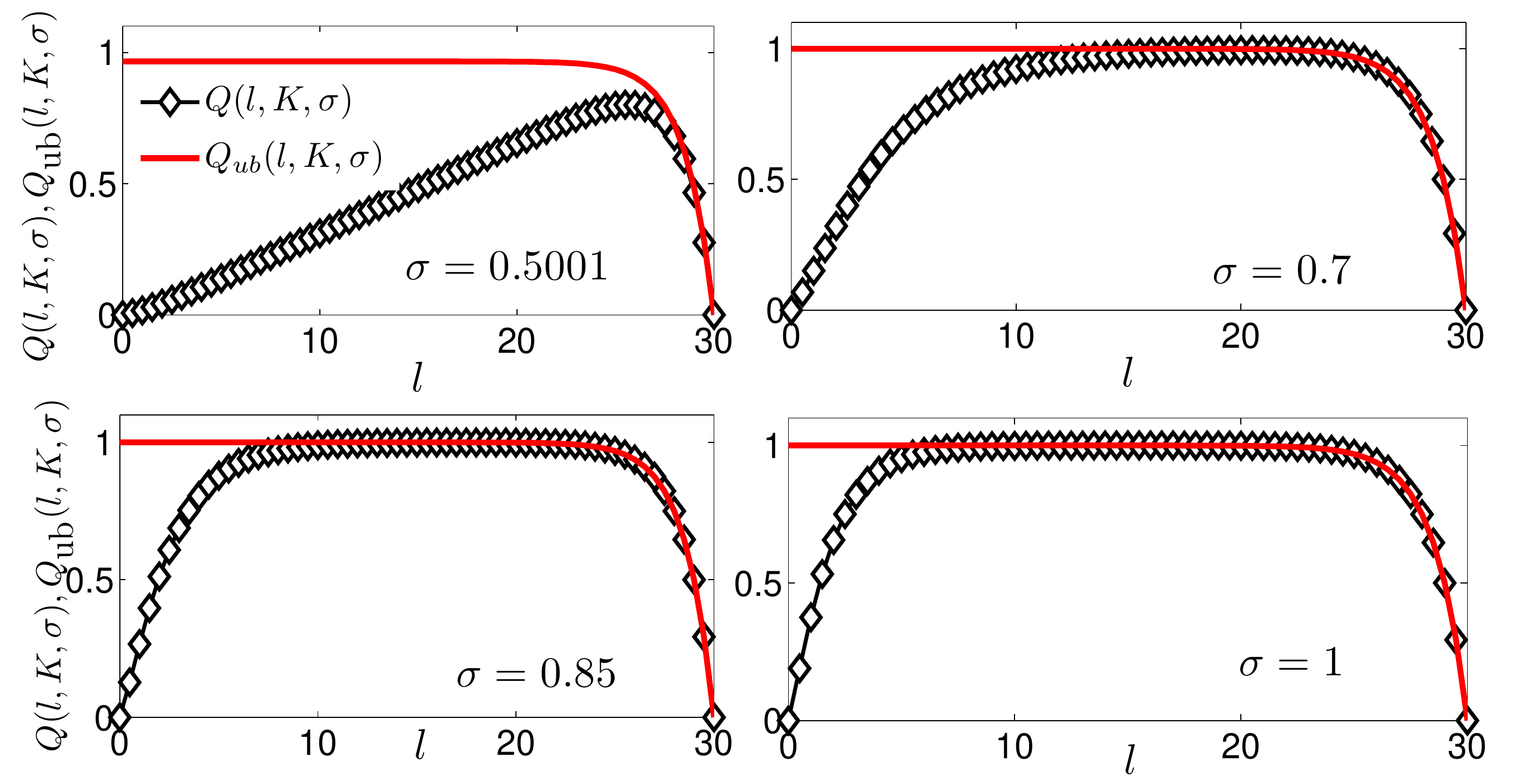}
	\caption{The modularity $Q(l, K, \sigma)$ ($\diamond$) and its upper bound $Q_{\textrm{ub}}(l, K, \sigma)$ (solid line) are evaluated from Eq.~\ref{eq:buono} and \ref{eq:modboundG} for a system of generation $K=30$ and plotted versus the logarithmic cluster size $l\in[0,30]$ for different values of $\sigma$. The function $Q(l, K, \sigma)$ shows a peak which moves from right to left as $\sigma$ grows. Moreover, as $\sigma$ grows there is a wider and wider region of values of $l$ where $Q(l, K, \sigma)$ is close to the maximum value and where the estimate provided by $Q_{\textrm{ub}}(l, K, \sigma)$ is good.}\label{modularity}
\end{figure}

In graph theory the modularity $Q$ is meant as a measure of the quality of a particular division of a network into a set of clusters (or groups, or communities) \cite{NewmanGirvan-PRE2004,Newman-PRE2004,Mieghen-2011}. More precisely, for an unweighted graph, the modularity is the fraction of links falling within clusters minus the expected fraction in a network where links are placed at random (conditional on the given cluster memberships and the degrees of vertices). Thus, if the number of links within a group is no better than random, the modularity is zero. On the other hand, a modularity approaching one is indicative of a strong community structure, that is, a dense intra-group and a sparse inter-group connection pattern.
Denoting with $A$ the (unweighted) adjacency matrix of the graph, with $z_i \equiv \sum_j A_{ij}$ the degree of a node $i$, and with $L_A \equiv 1/2 \sum_i z_i = 1/2 \sum_j \sum_i A_{ij}$ the number of edges in the graph, the modularity can be written as
\be \label{eq:mod1}
Q = \frac{1}{2L_A} \sum_{i=1}^{N} \sum_{j=1}^{N} \left( A_{ij} - \frac{z_i z_j}{2 L_A}  \right) \delta(\mathcal{C}_i, \mathcal{C}_j),
\ee
where $\mathcal{C}_i$ is the cluster to which node $i$ is assigned.
The previous expression can be extended to assess community divisions in weighted networks as
\be \label{eq:mod}
Q = \frac{1}{2L_J} \sum_{i=1}^{N} \sum_{j=1}^{N} \left( J_{ij} - \frac{w_i w_j}{2 L_J}  \right) \delta(\mathcal{C}_i, \mathcal{C}_j),
\ee
where $L_J \equiv 1/2 \sum_i w_i = 1/2 \sum_j \sum_i J_{ij}$ is the overall weight spread over all the edges. Of course, Eq.~\ref{eq:mod} recovers Eq.~\ref{eq:mod1} as long as weights collapse to the value $1$. 
\newline
Let us write Eq.~\ref{eq:mod} in a more convenient way \cite{Mieghen-2011}.
First, we notice that the terms to be summed up in Eq.~\ref{eq:mod} can be considered as elements of the $N \times N$ matrix $\mathbf{M}$ given by
\be
\mathbf{M} = \mathbf{J} - \frac{1}{L_J} \mathbf{w} \cdot \mathbf{w}^T.
\ee
Moreover, for a partitioning of the network into $c$ clusters one can introduce the $N \times c$ ``community matrix'' $\mathbf{S}$ defined as follows. 
As anticipated, the $c \in [1, N]$ disjoint, non-empty subsets are denoted by $\{ \mathcal{C}_1, \mathcal{C}_2, ..., \mathcal{C}_c\}$; clearly, $\sum_{k=1}^c  |\mathcal{C}_k| = N$. Then, the first column of $\mathbf{S}$ is a vector $(\mathbf{e}_{|\mathcal{C}_1|}, \mathbf{0}_{N - |\mathcal{C}_1|})^T$, where $\mathbf{e}_{k}$ and $\mathbf{0}_{k}$ are, respectively, the ones and zeros vector of length $k$; the second column of $\mathbf{S}$ is a vector $(\mathbf{0}_{|\mathcal{C}_1|},  \mathbf{e}_{|\mathcal{C}_2|}, \mathbf{0}_{N - |\mathcal{C}_1| - |\mathcal{C}_2|})^T$; similarly for the remaining $c-2$ columns. Notice that $\mathbf{S}^T \mathbf{S}$ is a diagonal matrix where the $k$-th diagonal term is $|\mathcal{C}_k|$ and that, for an arbitrary couple of nodes $(i,j)$, $\delta(\mathcal{C}_i, \mathcal{C}_j) = \sum_{k=1}^c S_{ik} S_{jk}$.
As a result, Eq.~\ref{eq:mod} can be recast as
\be \label{eq:mod2}
Q = \frac{1}{2L_J} \sum_{k=1}^c \sum_{i=1}^N \sum_{j=1}^N S_{ik} M_{ij} S_{jk} = \frac{\mathrm{Tr}(\mathbf{S}^T \mathbf{M} \mathbf{S})}{2L_J}.
\ee
A spectral upper bound $Q_{\textrm{ub}}$ for the modularity can be found as (see \cite{Mieghen-2011} for a complete derivation)
\be \label{eq:modbound}
Q \leq \frac{\eta_{\textrm{max}} N}{2 L_J} \left( 1 - \frac{1}{N^2} \sum_{k=1}^c |\mathcal{C}_k|^2 \right) \equiv Q_{\textrm{ub}}, 
\ee
where $\eta_{\textrm{max}}$ is the largest eigenvalue of $\mathbf{M}$.
\newline
Now, the spectra of the modularity matrix is strongly related with the spectra of the weight matrix $\mathbf{J}$ (or, in the case of unweighted graphs, of the adjacency matrix $\mathbf{A}$), see e.g., \cite{Mieghen-2011}.
In particular, for regular graphs where $w_i = w, \forall i$, the eigenvalues $\{\eta_i\}$ of the modularity matrix $\mathbf{M}$ are equal to the eigenvalues $\{\lambda_i \}$ of $\mathbf{J}$ (and, in turn, under a proper shift to eigenvalues $\{\varphi_i \}$ of $\mathbf{L}$), but the largest eigenvalue $\lambda_K$ is replaced by the eigenvalue $0$. 
This is proven rigorously for unweigthed graphs \cite{Mieghen-2011} and the proof can be straightforwardly extended to regular weighted graphs.
\newline
Specifically, for the graph $\mathcal{G}$ under study, recalling the results in Eqs.~\ref{eq:star}-\ref{eq:star2}, the spectrum of $\mathbf{M}$ reads as
\beas
\eta_i &=& \lambda_i = \frac{N^{-2\sigma}(2^{2\sigma}-2)-2\times 2^{i(1-2\sigma)}(2^{2\sigma}-1)+2^{2\sigma}}{(2^{2\sigma}-1)(2^{2\sigma}-2)}, ~~ i=0,\cdots,K-1, \text{ with algebraic multiplicity }\frac{N}{2^{i+1}},\\
\eta_K &=& 0  \text{ with algebraic multiplicity } 1.
\eeas
and, of course, $L_J = Nw/2$. By plugging $\eta_{\text{max}}=\eta_{K-1} = w - \varphi_{K-1}$ in Eq.~\ref{eq:modbound} we can estimate the modularity of a given partition $\{ \mathcal{ C}_k \}_{k=1,...,c}$. For instance, in the case of equipartition where there are $c$ clusters of size $N/c$ we get, no matter how clusters are chosen,
\be \label{eq:modboundG}
Q_{\textrm{ub}}(\{\mathcal{C}_k\}_{k=1,...,c} | \mathcal{C}_k =N/c, \forall k) =  \frac{ 4^{\sigma}-2+4^{\sigma}N(1-4^{\sigma})+4^{\sigma}N^{2\sigma} }{4^{\sigma}-2 + 2 N(1-4^{\sigma})+4^{\sigma}N^{2\sigma}}\left( 1 - \frac{1}{c}  \right). 
\ee
%
If we choose as graph partitioning the one suggested by the pattern of weights, namely the one with $c=N/2$ clusters (i.e., the set of dimers), or the one with $c=N/4$ clusters, ..., or the one with $c=2$ clusters (i.e., the two main branches of the graph), 
we can obtain an exact expression for (\ref{eq:mod2}), as explained hereafter. We consider the matrix $\mathbf{S}$ and, computing the trace of the matrix $\mathbf{S}^T \mathbf{M} \mathbf{S}$ for different choices of $l\in[1,K]$, we have
\begin{eqnarray}
Q(l,K,\sigma)&=&\frac{1}{Nw}\left [ N\sum_{m=1}^{l}2^{m-1}J(m)-2^l\sum_{m=1}^{K}2^{m-1}J(m)\right]\\
&=&\frac{1}{w}\sum_{m=1}^{l}2^{m-1}J(m)-\frac{2^l}{N}\nonumber\\
\label{eq:buono}
&=&\frac{N^{2\sigma}}{4^{l\sigma}}\frac{2^{l+1}-2-2^{l+1+2\sigma}+4^{\sigma}+4^{\sigma(l+1)}}{2N-2-2^{1+2\sigma}N+4^{\sigma}+(2N)^{2\sigma}}-\frac{2^l}{N} \\
\label{eq:a1}
&\approx& 2(t-1)2^{l(1 - 2\sigma)} + (1-2t)2^{-2l \sigma}  - 2^{l-K} + 1,
\end{eqnarray}
where, to lighten the notation, we wrote $J(d) = J(d, K, \sigma)$ and we simply highlighted the dependence on the logarithmic size $l$ of each cluster $l = \log_2 (N/c)$, $l\in[1,K]$; also, in the last passage we used $t=4^{-\sigma}$ and the approximation follows the assumption $N \gg 1$.
Notice that $Q(l,K,\sigma)$ is a concave function of $l$ with a peak at $l^*$ which can be estimated by solving the following (see Eq.~\ref{eq:a1})
\be \label{eq:root}
\frac{\partial }{\partial l} Q(l,K,\sigma) = 0 \Rightarrow 2(t-1)(1-2 \sigma) 2^{-2l \sigma} - 2 \sigma (1 - 2t) 2^{-2l \sigma - l } -2^{-K}=0.
\ee 
We can find the explicit solution for particular values of $\sigma$. For instance, setting $\sigma =1$ we get
\be \nonumber
l^*(K, \sigma=1)  = \log_2 \left  [  \frac{ \sqrt[3]{2} N + \sqrt[3]{(-2 + \sqrt{4 - 2 N})^2 N^2} } {\sqrt[3]{4(-2 + \sqrt{4 - 2 N}) N}} \right] = \frac{1}{2} (\log_2 3 -1 + K) -\frac{1}{3} \sqrt{ \frac{2}{3}} \frac{2^{-K/2}}{\log 2} - \frac{4}{27} \frac{2^{-K}}{\log 2} + \mathcal{O}(2^{-3K/2}), 
\ee
where the expansion holds for $K \gg 1$; more simply, we get that $l^*(K, \sigma) \sim K/2$, that is, the partition which maximizes the modularity when $\sigma = 1$ is the one where clusters have a size scaling as $\sqrt{N}$.
From Eq.~\ref{eq:root} one can also see that, by decreasing $\sigma$, the root $l^*$ grows.  
In Fig.~\ref{modularity} we compare $Q$ and $Q_{\textrm{ub}}$ for a network of generation $K=30$.
The latter provides a very good estimate for large values of $l$ (i.e., clusters of large sizes) and over a range which increases with $\sigma$.
 
 \section*{Conclusions \label{sec:conclusions}}
In this work we considered the hierarchical graph $\mathcal{G}$ introduced by Dyson and we calculated the exact spectrum of its Laplacian matrix $\mathbf{L}$. 
\newline
We recall that $\mathcal{G}$ is a weighted fully-connected graph, whose pattern of weights is ruled by the decay rate $\sigma \in (1/2, 1]$. A notion of metric can be defined in such a way that a couple of nodes $(i,j)$ at distance $d_{ij}$ are connected by a link associated to the weight $J_{ij} \sim 4^{- d_{ij} \sigma}$. {Thus, $\sigma$ tunes the inhomogeneity of the pattern of weights (when $\sigma \rightarrow 1/2$ the pattern is more homogeneous, while when $\sigma =1$ distant nodes are only weakly connected}.

Here, exploiting the deterministic recursivity through which $\mathcal{G}$ is built, we are able to derive explicitly the whole set of its Laplacian eigenvalues and eigenvectors. In this way, a large class of problems embedded in $\mathcal{G}$ can be addressed analytically. 
In fact, the Laplacian matrix (also called the admittance matrix or Kirchhoff matrix, see e.g., \cite{Biggs-1993,Mohar-DM1992}) is a discrete analog of the Laplacian operator in multivariable calculus and it naturally arises in the analysis of dynamic processes (e.g., random walks) occurring on the graph but also in the investigation of the dynamical properties of connected structures themselves (e.g., vibrational structures and the relaxation modes) \cite{Doyle-1984,Alavi-1991,Chung-1994,Mohar-1997}. Also, several topological features (e.g., the number of spanning trees, the graph partitioning) can be quantified or, at least, bounded, through the Laplacian eigenvalues.

As examples of applications we studied the random walk moving isotropically in $\mathcal{G}$, the continuous-time quantum walk moving in potential given by $\mathcal{G}$, and the relaxation times of a polymer whose structure is described by $\mathcal{G}$.
\newline
As for the simple random walk, we found that the average mean time $T_j$ to first reach a given node $j$ depends (super) linearly with the graph size $N$, the exact dependence being qualitatively affected by $\sigma$. In particular, when the pattern of weights is more inhomogeneous $T_j$ grows faster with $N$, as expected due to the unlikelihood to reach the farthest nodes.
\newline
As for the quantum walk, we proved that the long time average, measuring how spread the walker finally gets, is finite and lower bounded by $1/3$ even in the limit of large size. This value has to be compared with the equipartition limit $1/N$ holding for the classical case, suggesting that the coherent transport is not well performing on $\mathcal{G}$. Also, in the presence of absorbing traps, according to their arrangement, surviving stationary states can be established. 
\newline
Next, in the generalized Gaussian framework, we investigate the dynamics of polymers constituted by beads identified with the nodes of $\mathcal{G}$ and connected by elastic springs whose coupling is identified with the weights encoded by $\mathbf{J}$.
Using the Laplacian eigenvalues of $\mathcal{G}$ we estimated the mean displacement, the loss and the storage moduli under the stimulation of an external force. In particular, we found that in the intermediate time (or frequency) domain the response of the polymer is enhanced when the pattern of weights is more inhomogeneous. {In the case $\sigma =1$ we also recovered the expected scaling corresponding to a spectral dimension $d_s=2$.}

Finally, we exploit our results on the Laplacian spectra in order to derive information about the modularity of the graph itself. In particular, we find that when the graph size is large the modularity is maximum (or close to its maximum value) for partitions of size scaling as the square root of the overall graph size.

The knowledge of the exact spectrum for $\mathcal{G}$ paves the way to further investigations in the phenomenologies related to systems embedded in hierarchical, modular structures. These analysis look particularly interesting given the peculiar behaviours evidenced, e.g., in cognitive and neuroscience \cite{Moretti-Nature2013,Odor-SciRep2015, Gabrielli-2016}, and in bioinformatics \cite{Ma-BMC2016,Zhao-BMC2006,Ravasz-Science2002}. For instance, the existence of a wide range of relaxation time scales may shed light on the existence of a number of metastable states for the ferromagnetic Dyson model \cite{ABGGTT-PRL2015} and on the feasibility of parallel retrieval in Dyson associative networks \cite{ABGGTT-NN2015}.
 
 \section*{Acknowledgements}
\noindent
The authors are grateful to Adriano Barra for enlightening discussions and interactions.
 \newline
 Sapienza University of Roma and GNFM-INdAM (Progetto Giovani 2016) are acknowledged for financial support.


\section*{Author contribution statement}

E.A. designed the research, F.T. and E.A. performed the research, E.A. and F.T. write the manuscript.

\section*{Additional information}

The authors declare no competing financial interests.\\
Correspondence to agliari@mat.uniroma1.it or flavia.tavani@sbai.uniroma1.it

\end{document}